\documentclass[12pt]{article}
\pdfoutput=1
\usepackage[top=2cm,textwidth=16.6cm,textheight=23.4cm]{geometry} 
\usepackage{mhchem}
\usepackage{xcolor}
\usepackage{cite}
\usepackage{hyperref}
\hypersetup{colorlinks=true,linkcolor=red,citecolor=red,anchorcolor=black,urlcolor=magenta}
\usepackage[toc,page]{appendix}
\usepackage{mathtools}
\usepackage{amsfonts}
\usepackage{bbold}
\usepackage{textcomp}
\usepackage{amsmath, amsthm, amssymb, mathtools,empheq,latexsym,dsfont}
\usepackage{bbm}
\usepackage{mathrsfs}
\usepackage{slashed, simplewick}
\usepackage[utf8]{inputenc}
\usepackage{graphicx,placeins}
\usepackage{makeidx}
\usepackage[font=small,labelfont=bf]{caption}
\usepackage{nicefrac}
\usepackage{subfigure}
\usepackage{array, bigdelim,multirow,multicol}
\usepackage[integrals]{wasysym}
\usepackage{fancybox}
\usepackage{bm}
\usepackage{float}
\usepackage{rotating}
\usepackage{colortbl}
\usepackage{booktabs}
\usepackage{adjustbox}
\usepackage{arydshln}
\usepackage{doi}
\graphicspath{{immagini/}}

\definecolor{Gray}{gray}{0.92}
\definecolor{darkgreen}{rgb}{0.0, 0.8, 0.0}

\definecolor{amethyst}{rgb}{0.6, 0.4, 0.8}
\definecolor{antiquefuchsia}{rgb}{0.57, 0.36, 0.51}
\definecolor{airforceblue}{rgb}{0.36, 0.54, 0.66}
\definecolor{ao}{rgb}{0.0, 0.5, 0.0}
\definecolor{auburn}{rgb}{0.43, 0.21, 0.1}
\definecolor{brightcerulean}{rgb}{0.11, 0.67, 0.84}
\newcommand{\be}{\begin{equation}}
\newcommand{\ee}{\end{equation}}
\newcommand{\bea}{\begin{eqnarray}}
\newcommand{\eea}{\end{eqnarray}}

\newcommand{\eq}[1]{eq.~(\ref{eq:#1})}

\newcommand{\ord}[1]{\mathcal{O}\left( #1 \right)}

\def\r{\mathbf{r}}
\def\1{\mathbf{1}}
\def\op{\mathbf{1^\prime}}
\def\opp{\mathbf{1^{\prime\prime}}}
\def\2{\mathbf{2}}
\def\3{\mathbf{3}}

\DeclareMathOperator{\diag}{\texttt{diag}}
\DeclareMathOperator{\re}{\texttt{Re}}
\DeclareMathOperator{\im}{\texttt{Im}}
\DeclarePairedDelimiter{\abs}{\lvert}{\rvert}

\makeatletter
\renewcommand*{\@fnsymbol}[1]{\ensuremath{\ifcase#1\or *\or  \mathsection\or \ddagger\or
\dagger\or \mathparagraph\or \|\or **\or \dagger\dagger
\or \ddagger\ddagger \else\@ctrerr\fi}}
\g@addto@macro\bfseries{\boldmath}
\makeatother

\allowdisplaybreaks

\begin{document}
 \unitlength = 1mm

\begin{titlepage}

\begin{flushright}
{SISSA 05/2021/FISI\\
FTUV-21-0119.3119\\
IFIC/21-02}
\end{flushright}
\vspace{1.6cm}

\begin{center}
 {\Large\bf Modular Invariant Dynamics\\[0.2cm]
and Fermion Mass Hierarchies around $\tau=i$}\\[0.8cm]
\renewcommand*{\thefootnote}{\fnsymbol{footnote}}
Ferruccio~Feruglio$^{\,1}$%
\,\footnote{E-mail: {\tt feruglio@pd.infn.it}},
Valerio~Gherardi$^{\,2}$%
\,\footnote{E-mail: {\tt vgherard@sissa.it}}, 
Andrea~Romanino$^{\,2,3}$%
\,\footnote{E-mail: {\tt romanino@sissa.it}}
and
Arsenii~Titov$^{\,1,4}$%
\,\footnote{E-mail: {\tt titov@pd.infn.it}}\\
\vspace{0.8cm}
$^1${\small\it
Dipartimento di Fisica e Astronomia `G.~Galilei', Universit\`a di Padova\\
INFN, Sezione di Padova, Via Marzolo~8, I-35131 Padua, Italy}\\[0.2cm]
$^2${\small\it
SISSA, International School for Advanced Studies,\\
INFN, Sezione di Trieste, Via Bonomea 265, I-34136 Trieste, Italy}\\[0.2cm]
$^3${\small\it
ICTP, Strada Costiera 11, I-34151 Trieste, Italy}\\[0.2cm]
$^4${\small\it
Departament de F\'isica Te\`orica, Universitat de Val\`encia\\ 
IFIC, Universitat de Val\`encia-CSIC,
Dr.~Moliner 50, E-46100 Burjassot, Spain}
\end{center}
\vspace{0.8cm}
\begin{abstract}
\noindent
We discuss fermion mass hierarchies within modular invariant flavour models.
We analyse the neighbourhood of the self-dual point $\tau=i$, 
where modular invariant theories possess a residual $Z_4$ invariance. 
In this region the breaking of $Z_4$ can be fully described
by the spurion $\epsilon\approx\tau-i$, that flips its sign under $Z_4$. 
Degeneracies or vanishing eigenvalues of fermion mass matrices, 
forced by the $Z_4$ symmetry at $\tau=i$,
are removed by slightly deviating from the self-dual point. 
Relevant mass ratios are controlled by powers of $|\epsilon|$. 
We present examples where this mechanism is a key ingredient to
successfully implement an hierarchical spectrum in the lepton sector, even in the presence of a non-minimal K\"ahler potential.
\end{abstract}


\end{titlepage}
\setcounter{footnote}{0}
\setcounter{page}{2}

\section{Introduction}
Revealing the origin(s) of fermion masses 
and explaining the structure of quark and lepton mixing
are among the deepest long-standing questions in particle physics. 
Looking for an organising principle 
behind the observed patterns of fermion masses and mixing, 
flavour symmetries have been proposed and extensively studied in 
the last several decades (see ref.~\cite{Feruglio:2019ktm} for a recent review). 
In spite of a significant theoretical effort resulting in many models 
able to describe certain pieces of the flavour puzzle, 
there is arguably no fundamental theory of flavour. 
Still, the concept of symmetry is admittedly among the best tools 
we have to search for such a theory. 

While in the pure bottom-up approach flavour symmetries 
are introduced as a new ingredient, in the top-down perspective 
they may arise from a UV completion of the Standard Model (SM), 
perhaps string theory. Today a theory of flavour fully derived from
string theory represents a formidable unsolved problem, due to the huge number
of possible solutions, and we might be led to consider a more modest
approach where the large freedom related to a bottom-up procedure is mitigated by some guiding principle.
In particular, in ref.~\cite{Feruglio:2017spp}, 
modular invariance arising in many string compactifications 
has been proposed as a candidate for flavour symmetry in the lepton sector.

In the simplest case, modular invariance arises from the compactification of
a higher dimensional theory on a torus or an orbifold. Size and shape of the compact space are 
parametrised by a modulus $\tau$ living in the upper-half complex plane, up to modular transformations.
These can be interpreted as discrete gauge transformations, 
related to the redundancy of the description.
The low-energy effective theory, relevant to the known particle species, has to
obey modular invariance and Yukawa couplings become functions 
of $\tau$. The framework has a big conceptual advantage. 
In a generic bottom-up approach, realistic flavour symmetries require
an ad-hoc symmetry breaking sector, with Vacuum Expectation Values (VEVs) of scalar multiplets~---~the flavons~---~carefully tailored
in size and orientation. In minimal schemes based on modular invariance, 
flavons are not needed and the scalar sector can be completely replaced by the moduli space. 
Moreover the action of modular invariance in generation space 
occurs through a well-defined set of finite groups. Continuous groups and many discrete groups
are not allowed, thus reducing the arbitrariness of the construction.

The proposal has been accompanied by several significant developments and activity in model building 
(for a review, see \cite{Feruglio:2019ktm} and references therein). 
Recently, the proposed framework 
of modular invariant supersymmetric theories of flavour 
has been extended to incorporate (in a non-trivial way) 
several moduli~\cite{Ding:2020zxw}. 
The latter might be needed to describe different sectors of a theory, 
i.e.\ quarks, charged leptons and neutrinos.
Moreover, the role of modular invariance as flavour symmetry 
has been intensively investigated in the last two years
from a top-down perspective~\cite{Baur:2019kwi,Baur:2019iai,Kobayashi:2020uaj,Ishiguro:2020tmo}, 
resulting, in particular, in the concept of eclectic flavour symmetries~\cite{Nilles:2020nnc,Nilles:2020kgo,Nilles:2020tdp,Nilles:2020gvu,Baur:2020jwc}.
A top-down construction incorporating several moduli 
has been very recently formulated in refs.~\cite{Ishiguro:2020nuf,Baur:2020yjl}.

Despite the appealing aspects, realistic realisations of the framework have still to face
several difficulties. In the vast majority of phenomenologically viable modular invariant models 
constructed so far, the observed hierarchies, 
in particular those between the charged lepton masses, 
are achieved by tuning free parameters entering the superpotential. 
Moreover, even if the requirement of modular invariance represents a severe constraint
at the level of the Yukawa couplings of the (supersymmetric) theory, 
it allows in general a much larger freedom at the level of kinetic terms~\cite{Feruglio:2017spp,Chen:2019ewa}.
Non-minimal kinetic terms are allowed, with
corresponding free parameters that affect the prediction of fermion masses and mixing angles.
Finally, so far the modulus is mainly treated as a free parameter, varied to optimise the agreement
with the data. In concrete models, the preferred value of $\tau$ often occurs in the vicinity of the self-dual point $\tau=i$ (see, e.g.\ refs.~\cite{Feruglio:2017spp,Kobayashi:2018scp,Criado:2018thu,Novichkov:2018ovf,Novichkov:2019sqv,Okada:2019uoy,Criado:2019tzk,Kobayashi:2019gtp,Okada:2020rjb,Novichkov:2020eep,Wang:2020dbp,Okada:2020brs}),
where one of the generators of the modular group remains unbroken. In CP invariant realisations, it has been realised that CP violation can be explained
by a small departure of the modulus from this special point, where CP is unbroken~\cite{Feruglio:2017spp,Novichkov:2019sqv}. It remains to be understood how the
dynamics drives the modulus in the vicinity of, but not exactly on, the self-dual point. 
 
In this article, we aim to address the question of whether the
lepton mass hierarchies can originate from modular invariance alone, 
with the Lagrangian parameters being $\mathcal{O}(1)$ quantities. 
Some attempts to explain mass hierarchies avoiding fine tuning 
have been made in refs.~\cite{King:2020qaj,Okada:2020ukr}. 
In the former study, new fields singlet under both
the SM gauge group and finite modular group, 
but carrying non-trivial modular weights, have been introduced, 
whereas in the latter work expansions of modular forms 
around fixed points $\tau=i$, $\tau=\omega \equiv e^{2\pi i/3}$ and $\tau=i\infty$
have been employed, and a semi-analytic study of the fermion mass matrices 
has been performed~\footnote{Models of lepton masses and mixing angles with $\tau$ sitting exactly
at these fixed points have been discussed in refs.~\cite{Novichkov:2018ovf,Novichkov:2018nkm,Novichkov:2018yse,King:2019vhv,Gui-JunDing:2019wap,Gehrlein:2020jnr}.}.
We go beyond previous works discussing in detail 
the structure of the modular invariant theory
near $\tau = i$, which is motivated by phenomenological models.
 
At this point, the theory has a residual $Z_4$ symmetry 
(cf. ref.~\cite{Novichkov:2020eep}). 
We show that the action of this symmetry 
can be realised linearly, even when $\tau \neq i$. 
In particular, $\epsilon \approx \tau - i$ behaves as a (small) 
spurion with $Z_4$ charge $+2$. 
Thus, in a vicinity of $\tau = i$, we have 
a $Z_4$ symmetric theory broken by $\epsilon$. 
Similar considerations apply to other fixed points as well.
We show that the residual $Z_4$ symmetry can be exploited
to reduce the rank of the charged lepton mass matrix at $\tau = i$,
with small non-vanishing masses arising from a small departure
from the self-dual point.

As we will see in explicit models, when a minimal K\"ahler potential is adopted, the constraints 
coming from modular invariance realised near $\tau = i$
can be too restrictive to allow full agreement with the experimental data.
We are thus led to discuss non-minimal candidates of K\"ahler potential,
their pattern around $\tau = i$ and their impact on the predictions for masses, mixing angles 
and CP-violating (CPV) phases. 
The presence of extra inputs related to 
a more general K\"ahler potential is expected to reduce the predictability of our setup.
Nevertheless, 
our analysis suggests that
the arbitrariness coming from the K\"ahler potential can be partially tamed
precisely by mass matrices of reduced rank. Moreover, in the models
discussed in this paper, a non-minimal K\"ahler potential allows to obtain full agreement 
with the data, without being the dominant source of the observed hierarchies.
Further constraints on the K\"ahler potential and consequent
improvement in predictability could come by extending the modular group
to an eclectic flavour symmetry, where modular invariance is enhanced by a traditional finite flavour symmetry~\cite{Nilles:2020kgo}.

The article is organised as follows. 
In Section~\ref{S2}, we review the formalism of 
supersymmetric modular invariant flavour theories. 
Next, in Section~\ref{S3}, we zoom in on their structure near $\tau = i$ and
discuss linear realisation of the associated residual $Z_4$ symmetry. 
Further, in Section~\ref{S4}, we present examples of 
modular invariant models at level 3
where flavour hierarchies are generated by a small departure from $\tau = i$. 
Finally, we draw our conclusions in Section~\ref{S5}. 
Appendix~\ref{app:Kahler} discusses the most general 
form of the K\"ahler potential quadratic in the modular forms 
of level 3 and weight 2.

\section{Modular invariant models}
\label{S2}
In this section we shortly review the formalism of supersymmetric modular invariant theories~\cite{Ferrara:1989bc,Ferrara:1989qb} applied to flavour physics~\cite{Feruglio:2017spp}.
The theory depends on a set of chiral supermultiplets $\varphi$ comprising the dimensionless modulus $\tau\equiv\varphi_0/\Lambda$ (${\tt Im}~ \tau>0$) and other superfields $\varphi_i$~$(i\geq1)$. 
Here $\Lambda$ represents the cut-off of our effective theory, and can be interpreted as the relevant mass scale of an underlying fundamental theory.
In the case of rigid supersymmetry, 
the Lagrangian $\mathscr{L}$~\footnote{Up to terms with at most two derivatives in the bosonic fields.} is fully specified by the K\"ahler potential $K(\varphi,\bar{\varphi})$, a real gauge-invariant  function of the chiral multiplets and their conjugates, by the superpotential $W(\varphi)$, a holomorphic gauge-invariant  function of the chiral multiplets,
and by the gauge kinetic function $f(\varphi)$, a dimensionless holomorphic gauge-invariant function of the chiral superfields. Neglecting gauge interactions, we have:
\be
\label{lag}
\mathscr{L}=\int d^2\theta d^2\bar\theta~ K(\varphi,\bar{\varphi})+\int d^2\theta~ W(\varphi)+\int d^2\bar\theta~ {\overline W}(\bar\varphi)~~~.
\ee
The Lagrangian is invariant under transformations $\gamma$ of the homogeneous modular group $\Gamma=SL(2,Z)$:
\be
\label{mod}
\tau\to\gamma\tau=\frac{a \tau+b}{c\tau+d}~~~,~~~~~\varphi_i\to (c\tau+d)^{-k_i}\rho(\tilde\gamma)_{ij} \varphi_j~~~~(i,j\geq 1)~~~,
\ee
where $a$, $b$, $c$, $d$ are integers obeying $ad-bc=1$. Such transformations are generated by the two elements
of $\Gamma$:
\be
S=
\begin{pmatrix}
0&1\\
-1&0
\end{pmatrix} \qquad \text{and} \qquad
T=
\begin{pmatrix}
1&1\\
0&1
\end{pmatrix}~~.
\ee
The matrix $\rho(\tilde\gamma)$ is a unitary representation of 
the group $\Gamma_N=\Gamma/\Gamma(N)$, obtained as a quotient between the group $\Gamma$ and a principal congruence subgroup $\Gamma(N)$, the positive integer $N$ being the level of the representation. 
The level $N$ is kept fixed in the construction, and $\tilde{\gamma}$ 
represents the equivalence class of $\gamma$ in $\Gamma_N$.
In general $\rho(\tilde\gamma)$ is a reducible representation and all superfields
belonging to the same irreducible component should have the same weight $k_i$, here assumed to be integer
~\footnote{We restrict to integer modular weights. Fractional weights
are in general allowed, but require a suitable multiplier system \cite{Liu:2020msy,Nilles:2020nnc}.}.
In the following, we denote by $(\varphi_i, \psi_i)$ the spin-$(0, 1/2)$ components of the chiral superfields
$\varphi_i$ $(i \geq 1)$~%
\footnote{The distinction between superfields and their scalar components should be clear from the context.}.
The terms bilinear in the fermion fields read~\cite{Brignole:1996fn}:
\begin{equation}
\mathscr{L}_{F}=\mathscr{L}_{F, K}+\mathscr{L}_{F, 2}\,,
\end{equation}
with~%
\footnote{The covariant derivative is
$D_{\mu}\psi^{i}=\partial_{\mu}\psi^i+\left(K^{-1}\right)^i_mK^m_{kl}\partial_{\mu}\varphi^k
\psi^l$.
}:
\begin{equation}
\label{eq:Lag_modu}\mathscr{L}_{F, K}=i\,K^j_i\,\overline{\psi}_j\bar{\sigma}^{\mu}D_{\mu}\psi^{i}~,~~~~~
\mathscr{L}_{F, 2}=
-\frac{1}{2}\left[W_{ij} - W_l (K^{-1})^l_m
K^m_{ i j }\right]\psi^{i}\psi^{j}+\text{h.c.}~,
\end{equation}
where lower (upper) indices in $K$ and $W$ stand for derivatives with respect to holomorphic (anti-holomorphic) fields.
When the scalar fields in eq.~\eqref{eq:Lag_modu} take their VEVs, we can move to the basis where
matter fields are canonically normalised, through a transformation:
\be
\psi^{i}\to (z^{-1/2})_{ij}\psi^{j}~~~,
\ee
where the matrix $(z^{1/2})_{ij}$ satisfies: $K^j_i=[(z^{1/2})^\dagger]^{jl} (z^{1/2})_{li}$~%
\footnote{Notice that this
transformation mixes holomorphic and
anti-holomorphic indices, and there is no more fundamental distinction
between upper and lower components of the matrix $(z^{1/2})$.}. We can identify the fermion mass matrix as:
\be
\label{eq:fermion_mass_mat}
m_{kn}=\left[W_{ij} - W_l (K^{-1})^l_m
K^m_{ i j }\right](z^{-1/2})_{ik}(z^{-1/2})_{jn}~~~,
\ee
where VEVs are understood.
In the previous equation, the second term in the square bracket vanishes when supersymmetry is unbroken
and the VEV of $W_l$ is zero. When we turn on supersymmetry breaking effects, the first term 
is expected to dominate over the second one, provided there is a sufficient gap
between the sfermion masses $m_\mathrm{SUSY}$ and the messenger/cutoff scale $M$. This holds both for vector-like and for chiral fermions. Indeed, up to loop factors or other accidental factors,
the VEVs of $W_l$, $W_{ij}$ and $K^m_{ i j }$ are of the order of $m_\mathrm{SUSY}M$, $M$
and $1/M$, respectively, when fermions are vector-like. When chiral fermions are 
considered, $W_{ij}$ and $K^m_{ i j }$ are both depleted by $v/M$ with respect to the vector-like case, $v$ denoting the gauge symmetry breaking scale.
Thus we have a relative suppression between the two contributions of order $m_\mathrm{SUSY}/M$, which can be made tiny (cf. ref.~\cite{Criado:2018thu}). 
If we work under this assumption, the mass matrix is well approximated by:
\be
\label{mass}
m_{kn}=W_{ij}~(z^{-1/2})_{ik}(z^{-1/2})_{jn}~~~.
\ee
The supersymmetry breaking terms neglected here can be useful to give masses to light fermions, which otherwise would remain massless in the exact supersymmetry limit. We will come back to this point when discussing concrete models, in Section~\ref{S4}. 
Due to the conservation of the electric charge, the equality of eq.~\eqref{mass} holds separately in any charge sector. By focusing on the lepton sector $(E^c,L)$ and
by assuming that the neutrino masses arise from the Weinberg operator, we have:
\begin{equation}
\label{super}
W=-E^{c}_i\,\mathcal{Y}^{e}_{ij}(\tau)L_jH_d-\frac{1}{2\Lambda_L}L_i\,\mathcal{C}^{\nu}_{ij}(\tau)L_jH_uH_u~~~,
\end{equation}
where $H_{u,d}$ are the Higgs chiral multiplets and $\Lambda_L$ is the scale where lepton number is broken.
The general relation~\eqref{mass} specialises into:
\be
m_e=(z^{-1/2}_{E^c})^T{\mathcal Y}^{e}(\tau)(z^{-1/2}_L)~v_d~~~,~~~~~~m_\nu=(z^{-1/2}_L)^T{\mathcal C}^\nu(\tau)(z^{-1/2}_L)~v_u^2/\Lambda_L~~~,
\label{lepm}
\ee
where we have absorbed the renormalisation factors for $H_{u,d}$ in the definition of their VEVs.
In Section~\ref{S4}, we will also comment on the special limit where $z^{-1/2}_{E^c,L}$ are universal, i.e.\ proportional 
to the unit matrix. The mass matrices obtained in this case will be referred to as ``bare'' matrices and denoted by
$m_{e,\nu}^{(0)}$.
An important consequence of modular invariance is the special functional dependence of ${\mathcal Y}^{e}(\tau)$
and ${\mathcal C}^{\nu}(\tau)$ on the modulus $\tau$. Under a transformation of $\Gamma$, the chiral multiplets $(E^c_i,L_i,H_{u,d})$ transform as in eq.~\eqref{mod}, with weights $(k_{E^c_i},k_{L_i},k_{H_{u,d}})$ and representations
$(\rho_{E^c}(\tilde\gamma),\rho_{L}(\tilde\gamma),\mathbb{1})$.
For the superpotential $W$ to be modular invariant, 
${\mathcal Y}^{e}(\tau)$ and ${\mathcal C}^{\nu}(\tau)$ should obey:
\be
{\mathcal Y}^{e}(\gamma\tau)=(c\tau+d)^{{k_e}}~\rho_{E^c}^*(\tilde\gamma)~ {\mathcal Y}^{e}(\tau)
\rho_L^\dagger(\tilde\gamma)~,~~~~~
{\mathcal C}^{\nu}(\gamma\tau)=(c\tau+d)^{{k_\nu}}~\rho_L^*(\tilde\gamma)~ {\mathcal C}^{\nu}(\tau)
\rho_L^\dagger(\tilde\gamma)~,
\ee
where the weights $k_{e,\nu}$ are matrices satisfying: $(k_e)_{ij}=k_{E^c_i}+k_{L_j}+k_{H_d}$ and  $(k_\nu)_{ij}={k_L}_i +{k_L}_j+2k_{H_u}$. Thus ${\mathcal Y}^{e}(\tau)$ and ${\mathcal C}^{\nu}(\tau)$ are modular forms of given level
and weight. Since the linear space of such modular forms is finite dimensional, the choices for ${\mathcal Y}^{e}(\tau)$ and ${\mathcal C}^{\nu}(\tau)$ are limited. 
If neutrino masses originate from a type I seesaw mechanism, eqs.~\eqref{super}
and \eqref{lepm} hold with the identification: 
\be
\label{seesaw}
\frac{{\mathcal C}^{\nu}(\tau)}{\Lambda_L}=-({\mathcal Y}^{\nu}(\tau))^T~{\mathcal M}(\tau)^{-1}~
{\mathcal Y}^{\nu}(\tau)~~~,
\ee 
where ${\mathcal Y}^{\nu}(\tau)$ and ${\mathcal M}(\tau)$ denote the matrix of neutrino Yukawa couplings and the mass matrix of the heavy 
electroweak singlets $N^c$, respectively. Notice that there is no dependence on the renormalisation factor $(z^{-1/2}_{N^c})$ of the heavy modes.
In some cases ${\mathcal Y}^{e}(\tau)$ and/or ${\mathcal C}^{\nu}(\tau)$  are completely determined as a function of $\tau$
up to an overall constant, thus providing a strong potential constraint on the mass spectrum, eq.~\eqref{lepm}.

Unfortunately, such property does not extend to the K\"ahler potential $K$ and to the renormalisation factors
$(z^{-1/2}_{E^c,L})$. Minimal choices of $K$, appropriate for a perturbative regime, can receive large non-perturbative corrections in the region of the moduli space we will consider. Without a control over the non-perturbative dynamics, in a generic point of the moduli space the factors $(z^{-1/2}_{E^c,L})$ remain unknown. 
If we allowed for completely arbitrary $(z^{-1/2}_{E^c,L})$, under mild conditions any mass matrix could be predicted. From eq.~\eqref{lepm} we see that, given ${\mathcal Y}^{e}(\tau)$ and $(z^{-1/2}_L)$, we could reproduce any desired matrix $m_e$, by selecting a particular $(z^{-1/2}_{E^c})^T$:
\be
\label{arb}
(z^{-1/2}_{E^c})^T=m_e~({\mathcal Y}^{e}(\tau)v_d (z^{-1/2}_L))^{-1}~~~.
\ee
An arbitrary matrix $m_e$ would result in a completely unconstrained lepton mixing matrix.
Similar considerations would apply to the neutrino mass matrix $m_\nu$.

The loss of predictability associated to the K\"ahler corrections may however be less severe than eq.~\eqref{arb} might suggest, for two reasons. First, note that the above solution requires a non-singular ${\mathcal Y}^{e}(\tau)$. A singular ${\mathcal Y}^{e}(\tau)$ can only give rise to a singular $m_e$. Correspondingly, a hierarchical ${\mathcal Y}^{e}(\tau)$ can only correspond to a hierarchical $m_e$, unless the eigenvalues of the matrix $(z^{-1/2}_{E^c})^T$ in eq.~\eqref{arb} come in very large ratios. Although we cannot exclude the latter possibility, here we focus on the class of
models where the corrections associated to the K\"ahler potential do not alter the ``bare'' limit by more than about one order of magnitude.
Hence a singular or nearly singular ${\mathcal Y}^{e}(\tau)$ will tame the loss of predictability associated with the K\"ahler potential. Needless to say, a hierarchical ${\mathcal Y}^{e}(\tau)$ is needed to reproduce the mass spectrum in the charged lepton sector. Different considerations apply to the neutrino sector, where a singular ${\mathcal C}^\nu(\tau)$ might not be a good first order approximation of the data. 

A second constraint on the effect of the K\"ahler corrections arises in the vicinity of the fixed
points of $\Gamma$, $\tau=i$, $\tau=-1/2+i\sqrt{3}/2$ and $\tau=i\infty$, invariant under the action 
of the elements $S$, $ST$ and $T$, respectively. In the following, we will assume the modulus to be in the vicinity of the point $\tau = i$, as suggested by several models to correctly reproduce the data. The invariance under $S$ provides a constraint on the possible form of the K\"ahler potential at $\tau=i$ and in its vicinity.

\section{Residual symmetry near $\tau=i$}
\label{S3}
The residual symmetry of the theory at $\tau=i$ is the cyclic group $Z_4$ generated by the element $S$, whose action on the chiral multiplets $\varphi_i$ in $\tau=i$ can be read from eq.~\eqref{mod}:
\be
\varphi_i\to  \sigma_{ij}\varphi_j~~~,~~~~~~~~~~~~~~\sigma_{ij}=i^{k_i}\rho(\tilde S)_{ij}~~~~(i,j>0)~~~,
\label{eq:sigma}
\ee
where $\sigma$ is unitary, $\sigma^2$ is a parity operator and $\sigma^4=\mathbb{1}$. To analyse the neighbourhood of $\tau=i$, we expand both the K\"ahler potential and the superpotential in powers of the matter fields $\varphi_i$ $(i>0)$~%
\footnote{Electrically neutral multiplets whose scalar component acquires a VEV, like $H_{u,d}$, might mix
in the kinetic term with the modulus $\tau$. The mixing is parametrically suppressed by $v/\Lambda$ and will be ignored
in the following.}: 
\begin{equation}
\label{expan}
\begin{aligned}
W&=\sum_{i_1,...,i_n}Y_{i_1,...,i_n}(\tau)~\varphi_{i_1}...\varphi_{i_n}+...\\
K&=
\overline{\varphi}_i\, z^i_{~j}(\tau,\bar\tau)\,\varphi_j+...
\end{aligned}~~~.
\end{equation}
In the vicinity of $\tau=i$, it is possible to cast the theory as an ordinary $Z_4$ invariant theory, where the symmetry acts linearly on the fields, slightly broken by the spurion $(\tau-i)$. When we depart from $\tau=i$, the $S$ elements acts on the fields as:
\be
\label{es}
\tau\to-\frac{1}{\tau}~~~,~~~~~~
\varphi_i\to  (-\tau)^{-k_i}\rho(\tilde S)_{ij}\varphi_j~~~~(i,j>0)~~~.
\ee
We perform the field redefinition:
\be
\begin{cases}
\displaystyle \tau = i\, \frac{i + \epsilon/2}{i - \epsilon/2} \\[0.3cm]
\displaystyle \tilde{\varphi}_j= \left(1 -i \,\frac{\epsilon}{2}\right)^{-k_j} \varphi_j
\end{cases}
\ee
mapping the upper-half complex plane into the disk $|\epsilon| < 2$.
In the linear approximation: 
\be
\epsilon = (\tau - i) + O\left((\tau - i)^2\right)~~~.
\ee
Under the $S$ transformation in~\eqref{es}, the new fields transform as: 
\begin{equation}
\left\{
\begin{aligned}
& \epsilon \to-\epsilon \\[1mm]
& \tilde{\varphi}_i\to \sigma_{ij}~\tilde{\varphi}_j
\end{aligned}
\right.
\;.
\label{eq:linearized}
\end{equation}
We see that the action of the $Z_4$ symmetry is linear in the new field basis, even when $\tau\ne i$. In particular $\epsilon$ behaves as a spurion with $Z_4$ charge $+2$. In the new field basis, the coefficients of the field expansion (\ref{expan}) read:
\begin{equation}
\begin{aligned}
\tilde{Y}_{i_1,\dots,i_n} &=  
\left(1 - i\frac{\epsilon}{2}\right)^{k_{i_1}+\ldots+ k_{i_n}} 
Y_{i_1,\dots,i_n} \\
\tilde{z}^{i}_{~j} &= \left(1 + i \frac{\bar\epsilon}{2}\right)^{k_i} z^{i}_{~j} \left(1 -i \frac{\epsilon}{2}\right)^{k_j}
\end{aligned}~~~.
\end{equation}
The invariance of the theory under $Z_4$ requires $\tilde{Y}_{i_1,...,i_n}(\epsilon)$ and $\tilde{z}^i_{~j}(\epsilon,\bar\epsilon)$ to satisfy:
\begin{equation}
\begin{aligned}
\tilde{Y}_{i_1,...,i_n}(\epsilon)&=\sigma_{j_1 i_1}~...~\sigma_{j_n i_n}\tilde{Y}_{j_1,...,j_n}(-\epsilon) \\[0.2cm]
\tilde{z}^i_{~j}(\epsilon,\bar\epsilon)&=\sigma^{\dagger i k}~\tilde{z}^k_{~l}(-\epsilon,-\bar\epsilon)~\sigma_{l j}
\end{aligned}~~~.
\label{eq:Z4constraints}
\end{equation}
In particular, setting $\epsilon=0$, the above equations express the necessary conditions for the invariance
of the theory at the symmetric point $\tau=i$. By expanding $\tilde{z}^i_{~j}(\epsilon,\bar\epsilon)$ in powers of $\epsilon$ we see that the terms of first order vanish, up to possible non-diagonal terms relating
fields with opposite value of $\sigma$.
We conclude that in a neighbourhood of the fixed point $\tau=i$, and in the absence of any information
about the K\"ahler potential, the theory reduces to a linearly realised $Z_4$ flavour symmetric
theory, in the presence of a (small) spurion with charge $+2$.  

\section{Models}
\label{S4}
In this section, we present two models making use of the results of the previous section to account for the observed hierarchies in the lepton spectrum, namely the smallness of the charged
lepton mass ratios, $m_e/m_\tau$ and $m_\mu/m_\tau$ and of the neutrino mass ratio $r\equiv \delta m^2/|\Delta m^2|$, where $\delta m^2 \equiv m_2^2-m_1^2$ and $\Delta m^2 \equiv m_3^2 - (m_1^2+m_2^2)/2$ (with the standard neutrino labelling). The hierarchies will be naturally accounted for by the small breaking of $Z_4$, $|\epsilon| \ll 1$, i.e.\ by the closeness of the modulus $\tau$ to the $Z_4$ symmetric point $\tau = i$, while the parameters in the superpotential will be $\ord{1}$, and the corrections to the minimal K\"ahler will not be larger than $\ord{1}$.
In Table~\ref{tab:global_fit}, we collect
the best-fit values 
of the leptonic parameters with the corresponding 
$1\sigma$ 
uncertainties.
\begin{table}[t]
\centering
\renewcommand{\arraystretch}{1.5}
\begin{tabular}{|c|cc|} 
\hline
\texttt{Observable} & \multicolumn{2}{c|}{\texttt{Best-fit value with $1\sigma$ error}} \\ 
\hline
\hline
$m_e / m_\mu$ & \multicolumn{2}{c|}{$0.0048^{+0.0002}_{-0.0002}$} \\
$m_\mu / m_\tau$ & \multicolumn{2}{c|}{$0.0565^{+0.0045}_{-0.0045}$} \\ 
\hline
 & NO & IO \\
$\delta m^2\left[10^{-5}~\text{eV}^2\right]$ & \multicolumn{2}{c|}{$7.42^{+0.21}_{-0.20}$} \\
$\Delta m^2 \left[10^{-3}~\text{eV}^2\right]$ & $2.480^{+0.026}_{-0.028}$ & $-2.461^{+0.028}_{-0.028}$ \\
$r \equiv \delta m^2/|\Delta m^2|$ & $0.0299^{+0.0009}_{-0.0009}$ & $0.0301^{+0.0009}_{-0.0009}$ \\
$\sin^2\theta_{12}$ & $0.304^{+0.012}_{-0.012}$ & $0.304^{+0.013}_{-0.012}$ \\
$\sin^2\theta_{13}$ & $0.02219^{+0.00062}_{-0.00063}$ & $0.02238^{+0.00063}_{-0.00062}$ \\
$\sin^2\theta_{23}$ & $0.573^{+0.016}_{-0.020}$ & $0.575^{+0.016}_{-0.019}$ \\
$\delta/\pi$ & $1.09^{+0.15}_{-0.13}$  & $1.57^{+0.14}_{-0.17}$ \\
\hline
\end{tabular}
\caption{Best-fit values of the charged lepton mass ratios 
and the neutrino oscillation parameters 
with the corresponding 1$\sigma$ errors. 
For the charged lepton mass ratios we have used the values given in ref.~\cite{Ross:2007az}, 
averaged over $\tan\beta$ as described in the text,
whereas for the neutrino parameters we have used the results obtained in
refs.~\cite{Esteban:2020cvm,NuFITv50} 
(with Super-Kamiokande atmospheric data)}.
\label{tab:global_fit}
\end{table}
%
For the charged lepton mass ratios we use the results of ref.~\cite{Ross:2007az}, where for $m_\mu/m_\tau$ we take an average between 
the values obtained for $\tan\beta = 10$ and $\tan\beta = 38$. 
For the neutrino oscillation parameters we employ the results 
of the global analysis performed in refs.~\cite{Esteban:2020cvm,NuFITv50}.
In what follows, when fitting models to the data, we use 
five dimensionless observables that have been measured 
with a good precision, i.e.\ two mass ratios~%
\footnote{In the models presented below, $m_e = 0$ by construction, 
so we do not include the ratio $m_e/m_\mu$ here. 
See subsection~\ref{subsec:generating_me} for possible ways of generating non-zero $m_e$.}, 
$m_\mu/m_\tau$ and $r$,
and three leptonic mixing angles, 
$\sin^2\theta_{12}$, $\sin^2\theta_{13}$, $\sin^2\theta_{23}$.
Regarding the Dirac CPV phase, $\delta$, 
values between $\pi$ and $2\pi$ (approximately)
are currently allowed at $3\sigma$ for both neutrino mass spectrum 
with normal ordering~(NO) and that with inverted ordering~(IO). 
Moreover, in ref.~\cite{Novichkov:2018ovf}, it has been shown that under 
the transformation $\tau \to -\tau^\ast$ and complex conjugation of couplings 
present in the superpotential, CPV phases change their signs, 
whereas masses and mixing angles remain the same. 
In fact, this reflects CP properties of modular invariant models~\cite{Novichkov:2019sqv} (see also \cite{Kobayashi:2019uyt}).
As a consequence, the Dirac phase $\delta$ is not particularly constraining for our fits, and we do not include it in the list of input observables, 
regarding the obtained values as predictions.

\subsection{Model 1: Weinberg operator and inverted ordering}
\label{sec:model_1}
We work at level 3, and the relevant finite modular group is $\Gamma_3$. 
In this subsection, we assume that neutrino masses are generated 
by the Weinberg operator.
The field content of the model along with the assignment 
of $\Gamma_3$ representations and modular weights $k$
is shown in Table~\ref{tab:model_1}. The corresponding charges under $Z_4$, obtained using~\eq{sigma}, are shown in Table~\ref{tab:m1Z4}. We work in a real basis for the elements of $\Gamma_3$ where $\rho(\tilde{S})=\diag\,(+1,-1,-1)$ for the irreducible three-dimensional representation.

\begin{table}[t]
\centering
\renewcommand{\arraystretch}{1.5}
\begin{tabular}{|c|c|c|c|c|c|c|}
\hline
& $L$ & $E_{1}^{c}$ & $E_{2}^{c}$ & $E_{3}^{c}$ & $H_{u}$ & $H_{d}$ 
\\
\hline
\hline
$SU(2)_L\times U(1)_Y$& $(\2,-1/2)$& $(\1,+1)$& $(\1,+1)$& $(\1,+1)$& $(\2,+1/2)$& $(\2,-1/2)$ 
\\
\hline 
$\Gamma_{3}$ & $\3$ & $\1$ & $\1$ & $\op$ & $\1$ & $\1$ 
\\
\hline 
$k$ & $1$ & $3$ & $3$ & $3$ & $0$ & $0$ 
\\ 
\hline 
\end{tabular}
\caption{Assignment of 
representations and modular weights in Model 1.}
\label{tab:model_1}
\end{table}
%
\begin{table}[t]
\centering
\renewcommand{\arraystretch}{1.5}
\begin{tabular}{|c|c|c|c|c|c|c|c|c|c|}
\hline
& $\tilde{L}_1$ & $\tilde{L}_2$ & $\tilde{L}_3$ & $\tilde{E}_{1}^{c}$ & $\tilde{E}_{2}^{c}$ & $\tilde{E}_{3}^{c}$ & $\tilde{H}_{u}$ & $\tilde{H}_{d}$ & $\epsilon$
\\
\hline
\hline
$Z_4$ & $1$ & $-1$ & $-1$ & $-1$ & $-1$ & $-1$ & $0$ & $0$ & 2 
\\ 
\hline 
\end{tabular}
\caption{$Z_4$ charges (\texttt{mod} 4) in Model 1.}
\label{tab:m1Z4}
\end{table}
The quantum number assignments have immediate consequences 
for the
charged lepton mass spectrum:
\begin{enumerate}
\item At $\tau=i$, the charged lepton mass matrix $m_e$ has rank one. 
This follows from the $Z_4$ charges in Table~\ref{tab:m1Z4}, forcing
\be
\label{me1}
m_e=\begin{pmatrix}\alpha & 0 & 0\\
\beta & 0 & 0\\
\gamma & 0 & 0
\end{pmatrix}.
\ee
\item For a generic $\tau\neq i$, $m_e$ has rank two. While $Z_4$ alone would allow $m_e$ to have rank three, the underlying modular invariance forces the coefficients of the first and second rows of $m_e$ to be proportional, thus reducing the rank. In fact, modular invariance requires the coupling of
$E_{1}^{c}$ and $E_{2}^{c}$ to $L$ to be proportional to the same
modular form multiplet, namely, the triplet of weight four. The K\"ahler corrections cannot modify the rank condition. Thus, in the considered model, the electron has zero mass. 
\end{enumerate}
%
For $\tau \approx i+\epsilon$, with $\left|\epsilon\right|\ll1$,
we obtain the prediction
\begin{equation}
m_{e}\,\colon\,m_{\mu}\,\colon\,m_\tau=0\,\colon\,\mathcal{O}(\epsilon)\,\colon\,1\,.
\label{eq:Ratios mE - model 1}
\end{equation}
%
Concerning the neutrino mass spectrum, from the charges of the lepton doublets $L_i$ in Table~\ref{tab:m1Z4},
we deduce that $m_{\nu}$ in $\tau=i$ takes 
the following general form:
\begin{equation}
m_{\nu}=\begin{pmatrix}0 & a & b\\
a & 0 & 0\\
b & 0 & 0
\end{pmatrix}.
\label{eq:General neutrino Weinberg}
\end{equation}
This matrix has rank two and two
degenerate non-zero eigenvalues. 
Notice that, while a generic $Z_4$ model would not account for the values of the parameters $a$ and $b$, here the underlying modular invariance fixes the relative values, before K\"ahler corrections.  With the $Z_4$ assignment of Table \ref{tab:m1Z4}, we are implicitly using the basis where $S$ is diagonal for the irreducible triplet of $\Gamma_3$ and we find $a/b=Y_3(i)/Y_2(i)$, where $Y^{(2)}(\tau) \equiv (Y_1(\tau),Y_2(\tau),Y_3(\tau))^T$ denotes the weight-two triplet of modular forms. 
On the other hand, generic K\"ahler corrections could mix $L_2$ and $L_3$, as they have the same $Z_4$ charge (see \eq{Z4constraints}), leading to arbitrary $a/b$, as in generic $Z_4$ models. 
For $\tau\approx i+\epsilon$, the rank of $m_\nu$ becomes three, and we
obtain the neutrino mass spectrum with inverted ordering of the form
\begin{equation}
m_{1}\,\colon\,m_{2}\,\colon\,m_{3}=
1\,\colon \left(1+\mathcal{O}(\epsilon)\right) \colon\,\mathcal{O}(\epsilon)\,,
\label{eq:ratios mN - model 1}
\end{equation}
%
and, in particular, 
\begin{equation}
r=\mathcal{O}(\epsilon)\,.
\label{eq:ratio r - model 1}
\end{equation}
%
Clearly, both qualitative relations \eqref{eq:Ratios mE - model 1}
and \eqref{eq:ratio r - model 1} are phenomenologically intriguing.
They are consequences of modular invariance alone, and are thus independent
from the parameters in the superpotential or the K\"ahler potential 
(provided these are non-hierarchical by themselves). In subsection~\ref{subsec:generating_me}, we discuss two possible mechanisms to generate a naturally small electron mass. 

While this model successfully accounts for the observed mass hierarchies (with a non-vanishing electron mass still to be generated), it is not satisfactory when it comes to the mixing angles. The point is that in order for \eq{General neutrino Weinberg} to lead to a reasonable leading order approximation, the tau lepton should correspond to a linear combination of $L_2$ and $L_3$, while eq.~\eqref{me1} forces the tau lepton to be mainly $L_1$. Indeed, the prediction for the mixing angles at $\tau = i$ is
\begin{equation}
 \sin^2\theta_{13} = \cos^2\theta^e_{12}\,, \qquad
 \sin^2\theta_{12} = \frac{1}{2}\,, \qquad
 \sin^2\theta_{23} = 1\,,
 \label{eq:anglesLO}
\end{equation}
where $\theta^e_{12}$ is an arbitrary angle related to the presence of two vanishing eigenvalues in $m_e$, to be fixed by the $Z_4$ breaking. 
These predictions imply that in order to generate the correct mixing angles $\sin^2\theta_{23}\approx0.6$ and $\sin^2\theta_{12}\approx0.3$, large hierarchical deviations
from the minimal K\"ahler metrics are required~\footnote{The need for non-minimal K\"ahler metrics stems not only from 
the leading order predictions for the mixing angles, 
but also from the mass spectrum of the model. In the vicinity of $\tau=i$ we found, both numerically and through an
approximate analytical study, that $m_\mu/m_\tau$ is smaller than $r$, while data require the opposite.}, 
as $|\epsilon| \ll 1$ cannot give rise to such large corrections.
This is clearly an unpleasant feature, since it introduces a source of hierarchy in the Lagrangian parameters.
We carried out a full numerical study of the model, after adding a non-minimal K\"ahler potential depending
on four new real parameters. 
The outcome confirms the above qualitative considerations.
More precisely, we gauge the degree of hierarchy related to a non-canonical K\"ahler potential $K$ by means of the 
condition number
\be
\label{condnum}
\kappa(K) = \lambda_{\text{max}}(z)/\lambda_{\text{min}}(z)~~~,
\ee
the ratio between the maximum and minimum eigenvalues of $z^i_{~j}$ at the best-fit point. We find that all K\"ahler metrics providing a good fit near $\tau=i$ turn out to have $\kappa(K_{L,\,E^c})$ very large, typically in the range $10^3\div 10^4$.
We discuss in the next subsection a seesaw variant of the present model which allows to mitigate the need of hierarchical K\"ahler metrics. 

\subsection{Model 2: seesaw mechanism and normal ordering}
The main phenomenological obstructions in the model discussed above
are the leading order predictions for the mixing angles. In this subsection,
we show how to evade them by introducing 
electroweak singlet
neutrinos $N^{c}$ and generating the Weinberg operator through the 
type I seesaw mechanism. This widens the class of possible neutrino mass matrices that can be obtained, if the singlet neutrino mass matrix becomes singular in the limit $\tau\to i$. In this case, for the standard analysis of the seesaw mechanism to be valid, singlet neutrino masses are required to be large compared to the electroweak scale. In the example discussed below, this is easily achieved outside of a neighbourhood of $\tau = i$, provided the overall singlet neutrino mass scale is large enough.

To be concrete, we augment the field content of Table~\ref{tab:model_1}
with electroweak singlets $N^{c}\sim\3$ under $\Gamma_{3}$,
with weight $k_{N^c}=1$. As before, we denote by $Y^{(2)}$ the weight $2$ modular form triplet,
and by $Y^{(4)}\equiv(Y^{(2)} Y^{(2)})_{\3_S}$ 
the weight 4 triplet of modular forms. 
We denote by $\3_S$ and $\3_A$ the symmetric
and antisymmetric triplet contractions of two $\Gamma_{3}$ triplets,
respectively.
The superpotential $W=W_e+W_\nu$ of the lepton sector reads:
\begin{align}
W_{e} & =-
\left[\alpha E_{1}^{c} \left(LY^{(4)}\right)_{\1}
+ \beta E_{2}^{c} \left(LY^{(4)}\right)_{\1}
+ \gamma E_{3}^{c} \left(LY^{(4)}\right)_{\opp}\right]H_{d}\,,
\label{eq:Electron superpotential}\\[0.2cm]
W_{\nu} &=-\kappa\,\left(\left[\left(N^{c}L\right)_{\3_S}
+g\left(N^{c} L\right)_{\3_A}\right] Y^{(2)}\right)_{\1}H_{u}
-\Lambda\left(N^{c}N^{c}Y^{(2)}\right)_{\1}\,.
\label{eq:See saw superpotential}
\end{align}
The parameters $\kappa$ and $\Lambda$ can be made real 
without loss of generality, whereas $g$ is complex in general. 
In the real basis for $\3$ of $\Gamma_3$, this superpotential 
leads to the following matrices ${\mathcal Y}^{e}(\tau)$, $\mathcal{Y}^\nu(\tau)$ and $\mathcal{M}(\tau)$:
\begin{align}
{\mathcal Y}^{e}(\tau) & =2\begin{pmatrix}
\alpha Y_{2}Y_{3} & \alpha Y_{1}Y_{3} & \alpha Y_{1}Y_{2}\\
\beta Y_{2}Y_{3} & \beta Y_{1}Y_{3} & \beta Y_{1}Y_{2}\\
\gamma Y_{2}Y_{3} & \gamma \omega Y_{1}Y_{3} &  \gamma \omega^2 Y_{1}Y_{2}
\end{pmatrix},
\label{eq:bare_me}\\[2mm]
\mathcal{Y}^\nu(\tau)& = \kappa \left[
\begin{pmatrix}
0 & Y_3 & Y_2 \\
Y_3 & 0 & Y_1 \\
Y_2 & Y_1 & 0 
\end{pmatrix}
+ g \begin{pmatrix}
0 & Y_3 & -Y_2 \\
-Y_3 & 0 & Y_1 \\
Y_2 & -Y_1 & 0
\end{pmatrix}
\right], 
\label{eq:Ynu0}\\[2mm]
\mathcal{M}(\tau)& = 2 \Lambda \begin{pmatrix}
0 & Y_3 & Y_2 \\
Y_3 & 0 & Y_1 \\
Y_2 & Y_1 & 0
\end{pmatrix}.
\label{eq:M0}
\end{align}
%
The matrix $\mathcal{C}^\nu(\tau)$ of eq. (\ref{super}) is now given by the seesaw formula of eq.~\eqref{seesaw}.

Some analytical considerations easily follow from the previous equations for the ``bare'' quantities,
i.e.\ those corresponding to the minimal K\"ahler potential.
We make use of the following $\epsilon$-expansion 
of $Y^{(2)}~%
$\footnote{One can prove, in general, that $\left.\frac{\text{d}}{\text{d}\tau}\right|_{i}Y_{2,3}=i\,Y_{2,3}(i)$.
Moreover, we can rephase $Y^{(2)}$ in such a way that $(Y_{3}(i))^{*}=Y_{2}(i)\equiv y$.
In this basis, we find that $\left.\frac{\text{d}}{\text{d}\tau}\right|_{i}Y_{1}\in-i\,\mathbb{R}^{+}$.
}:
\begin{equation}
Y_{1} = -i x\epsilon\,, \qquad 
Y_{2} = y\left(1+i\epsilon\right)\,,\qquad 
Y_{3} = y^\ast\left(1+i\epsilon\right),
\label{eq:eps-expansion}
\end{equation}
%
where, up to an overall constant, $x\approx1.49087$ and 
$y=\sqrt{3}/2 + i (3/2-\sqrt{3})$.
To first order in $\epsilon$ we obtain:
\begin{equation}
\frac{\mathcal{Y}_\nu(\tau)}{\kappa} = \begin{pmatrix}
0 & \left(1+g\right) y^\ast & \left(1-g\right) y \\
\left(1-g\right) y^\ast & 0 & 0 \\
\left(1+g\right) y & 0 & 0 
\end{pmatrix}
+ i\,\epsilon \begin{pmatrix}
0 & \left(1+g\right) y^\ast & \left(1-g\right) y \\
\left(1-g\right) y^\ast & 0 & -\left(1+g\right) x \\
\left(1+g\right) y & -\left(1-g\right) x & 0
\end{pmatrix}, 
\label{eq:eps-expansion_Ynu0} 
\end{equation}
%
\begin{equation}
\frac{\mathcal{M}(\tau)}{2\Lambda} = \begin{pmatrix}
0 & y^\ast & y \\
y^\ast & 0 & 0 \\
y & 0 & 0
\end{pmatrix}
+i\,\epsilon \begin{pmatrix}
0 & y^\ast & y \\
y^\ast & 0 & - x \\
y & -x & 0
\end{pmatrix}.
\label{eq:eps-expansion_M0}
\end{equation}
%
Notice that the bare Majorana mass matrix has one eigenvalue proportional to $\epsilon$,
thus vanishing in the limit $\tau\to i$. 
This corresponds to the case of single right-handed neutrino dominance, 
in which one of the electroweak singlet neutrinos is massless in the symmetric limit~\footnote{As already observed, in order for the standard seesaw analysis to be valid, we must require the product $\abs{\epsilon} \Lambda$ (that is the order of magnitude of the lightest right-handed neutrino mass) to be large with respect to the electroweak scale. For the present model this does not pose any practical restriction, since the best-fit region (see below) is achieved for values of $\vert \epsilon \vert \sim 10^{-2} \div 10^{-1}$.
}. 
Inverting the matrix in eq.~\eqref{eq:eps-expansion_M0} and using the seesaw relation for the bare light neutrino mass matrix $m_\nu^{(0)}$ we find to $\mathcal{O}(\epsilon)$:
\begin{equation}
m_\nu^{(0)} = \begin{pmatrix}
\frac{2 i g^2 |y|^2}{x} \frac{1}{\epsilon} - \frac{4g^2|y|^2}{x} & -\left(1+g^2\right) y^\ast & -\left(1+g^2\right) y \\
-\left(1+g^2\right) y^\ast & 0 & 0 \\
-\left(1+g^2\right) y & 0 & 0
\end{pmatrix} 
+ \mathcal{O}(\epsilon)\,.
\label{eq:seesaw at LO}
\end{equation}
%
The leading order form of the charged lepton mass matrix is as in eq.~(\ref{me1}). 

The leading order predictions for neutrino masses and mixing angles strongly depend upon the parameter $g$. 
\begin{itemize}
\item
A neutrino mass spectrum with IO can be realised when $|g|^2 \ll |\epsilon|$. Then, $m_\nu^{(0)}$ has approximately the same form as in the model with the Weinberg operator considered in subsection~\ref{sec:model_1}.
We get the neutrino mass spectrum with IO:
\begin{equation}
m_{1}\,\colon\,m_{2}\,\colon\,m_{3}=
1\,\colon \approx 1 \colon\,\mathcal{O}(\epsilon)
\label{eq:neutrino_masses_g2lleps}
\end{equation}
%
and the predictions for the mixing angles reported in eq.~\eqref{eq:anglesLO}, 
in particular, $\sin^2\theta_{23} = 1$.
\item A neutrino mass spectrum with NO can be realised when $|\epsilon| \ll |g|^2$. 
In this case, 
\begin{equation}
m_\nu^{(0)} = \begin{pmatrix}
c(g^2/\epsilon) & a & b \\
a & 0 & 0 \\
b & 0 & 0
\end{pmatrix}+ \mathcal{O}(\epsilon)\,,
\end{equation}
%
with $|a|$, $|b|$, $|c|$ being $\mathcal{O}(1)$ numbers. Therefore, we have the neutrino mass spectrum with NO: 
\begin{equation}
m_{1}\,\colon\,m_{2}\,\colon\,m_{3}=
\mathcal{O}\left(\epsilon^2/g^2\right)\,\colon\,\mathcal{O}\left(\epsilon^2/g^4\right)\,\colon\,\mathcal{O}(1) 
\end{equation}
%
implying $r = O(\epsilon^4/g^4)$. 
The mixing angles are: 
\begin{equation}
 \sin^2\theta_{12} = \ord{1}\,, \qquad
 \sin^2\theta_{23} \approx \sin^2\theta_{13} \approx r^{1/2} \,. %
 \label{eq:anglesLO1}
\end{equation}
%

Again, the leading order prediction for $\sin^2\theta_{23}$ is far away from its measured value and requires significant corrections from the K\"ahler potential. 
\end{itemize}

To verify the viability of the model we have performed a full numerical study,
also allowing for a non-minimal form of the K\"ahler potential for the matter fields.
In general, modular invariance allows many terms in the K\"ahler potential~\cite{Feruglio:2017spp,Chen:2019ewa}. 
In the considered bottom-up approach, there seems to be no way of 
reducing the number of these terms. However, this may change if 
modular symmetry is augmented by a traditional finite flavour symmetry~\cite{Nilles:2020kgo} or, perhaps, if some other top-down principle is in action. In what follows, to be concrete, we adopt three simplifying assumptions: 
\begin{itemize}
 \item The new terms in $K$ are quadratic in $Y^{(2)}$. This is sufficient to illustrate our results.
 \item The minimal form (up to overall normalisation) is restored at $\im\tau \to \infty$. This assumption is inspired by the minimal form of the K\"ahler potential arising in 
certain string theory compactifications in the  
large volume limit, corresponding to $\im\tau \to \infty$ 
(see, e.g.\ \cite{Chen:2019ewa,Asaka:2020tmo} and references therein).
 \item The diagonal entries, already controlled
by the minimal K\"ahler potential, are not affected by the new terms.
\end{itemize}
%
Under these assumptions and with the assignment of 
representations and weights given in Table~\ref{tab:model_1}, 
we find~%
\footnote{We present the full expressions for $K_L$ 
and $K_{E^c}$ quadratic in $Y^{(2)}$ in Appendix~\ref{app:Kahler}.}:
\begin{equation}
K=L^{\dagger}K_{L}L+E^{c\dagger}K_{E^c}E^{c}\,,
\label{eq:Kahler potential}
\end{equation}
where
\begin{equation}
K_L = \frac{1}{2\im\tau} \begin{pmatrix}
1 & 0 & 0 \\
0 & 1 & 0 \\
0 & 0 & 1
\end{pmatrix} 
+ 2 \im\tau \begin{pmatrix}
0 & \left(\alpha_5+i\alpha_6\right)X_{12}  & \left(\alpha_5-i\alpha_6\right)X_{13} \\
\left(\alpha_5-i\alpha_6\right)X_{12}^\ast & 0 & \left(\alpha_5+i\alpha_6\right)X_{23}  \\
\left(\alpha_5+i\alpha_6\right)X_{13}^\ast & \left(\alpha_5-i\alpha_6\right)X_{23}^\ast & 0
\end{pmatrix}.
\label{eq:KL_two_param}
\end{equation}
%
Here $\alpha_5$ and $\alpha_6$ are real coefficients and
\begin{equation}
X_{12} = Y_1^\ast Y_2 - Y_1 Y_2^\ast\,,
\qquad
X_{13} = Y_1^\ast Y_3 - Y_1 Y_3^\ast\,,
\qquad
X_{23} = Y_2^\ast Y_3 - Y_2 Y_3^\ast\,.
\label{eq:Xij}
\end{equation}
%
In the $E^c$ sector, we obtain:
\begin{equation}
K_{E^c} = \frac{1}{8 \left(\im\tau\right)^3} \begin{pmatrix}
1 & 0 & 0 \\
0 & 1 & 0 \\
0 & 0 & 1
\end{pmatrix} 
+ \frac{1}{2\im\tau} \begin{pmatrix}
0 & 0  & c_{13} X \\
0 & 0 & c_{23} X \\
c_{13}^\ast X^\ast & c_{23}^\ast X^\ast & 0
\end{pmatrix},
\label{eq:KEc_two_param}
\end{equation}
%
where $c_{13}$ and $c_{23}$ are complex coefficients and
\begin{equation}
X = Y_1^\ast Y_1 + \omega Y_2^\ast Y_2 + \omega^2 Y_3^\ast Y_3 \sim \opp\,.
\label{eq:X}
\end{equation}
%
Noteworthy, the seesaw formula~\eqref{seesaw}
does not depend on the renormalisation factor $z_{N^c}^{-1/2}$ of the heavy
fields $N^{c}$, so that we will not need to specify the K\"ahler metric
of $N^{c}$ in what follows.
\begin{table}[t]
\footnotesize{
\centering
\renewcommand{\arraystretch}{1.5}
\begin{tabular}[t]{|c|c|}
\hline
\multicolumn{2}{|c|}{\texttt{Input parameters}} \\
\hline
\hline
$\re\tau$ & $\pm0.0235$ \\
$\im\tau$ & $1.080$ \\
$\beta/\alpha$ & $0.1459$ \\
$\gamma/\alpha$ & $5.955$ \\ 
$\re g$ & $-0.1494$ \\
$\im g$ & $\mp 0.3169$ \\
$\alpha_5$ & $-0.2071$ \\
$\alpha_6$ & $-0.1437$ \\
$c_{13}$ & $-0.2656$ \\
$c_{23}$ & $0.0145$ \\
\hdashline
$v_u^2 \kappa^2/\Lambda$~[eV] & $0.0189$ \\
\hdashline
$|\epsilon|\approx|\tau-i|$ & $0.0830$ \\
\hline
\end{tabular}
\hspace{0.2cm}
\begin{tabular}[t]{|c|c|}
\hline
\multicolumn{2}{|c|}{\texttt{Observables}} \\
\hline
\hline
$m_e/m_\mu$ & $0$ \\
$m_\mu/m_\tau$ & $0.0565$ \\
$r$ & $0.0299$\\
$\sin^2\theta_{12}$ & $0.304$ \\
$\sin^2\theta_{13}$ & $0.02219$ \\
$\sin^2\theta_{23}$ & $0.573$ \\
\hdashline
$\delta m^2~[10^{-5}~\text{eV}^2]$ & $7.42$ \\
$\Delta m^2~[10^{-3}~\text{eV}^2]$ & $2.480$ \\
\hline
\end{tabular}
\hspace{0.2cm}
\begin{tabular}[t]{|c|c|}
\hline
\multicolumn{2}{|c|}{\texttt{Predictions}} \\
\hline
\hline
$m_1$~[eV] & $0.0062$ \\
$m_2$~[eV] & $0.0106$ \\
$m_3$~[eV] & $0.0506$\\
$\delta/\pi$ & $\pm0.92$ \\
$\alpha_{21}/\pi$ & $\pm0.97$ \\
$\alpha_{31}/\pi$ & $\pm0.93$ \\
\hdashline
$|m_{ee}|$~[eV] & 0 \\
$\sum_i m_i$~[eV] & 0.0673 \\
\hdashline
Ordering & NO \\
\hdashline
$M_1/\Lambda$ & 0.225 \\
$M_2/\Lambda$ & 2.298 \\
$M_3/\Lambda$ & 2.524 \\
\hline
\end{tabular}
\caption{First pair of best-fit points in a vicinity of $\tau = i$ found considering the K\"ahler potential in eqs.~\eqref{eq:KL_two_param}--\eqref{eq:X}.}
\label{tab:bfp_model_2}
}
\end{table}
\begin{table}[t]
\footnotesize{
\centering
\renewcommand{\arraystretch}{1.5}
\begin{tabular}[t]{|c|c|}
\hline
\multicolumn{2}{|c|}{\texttt{Input parameters}} \\
\hline
\hline
$\re\tau$ & $\pm0.0328$ \\
$\im\tau$ & $1.137$ \\
$\beta/\alpha$ & $0.2388$ \\
$\gamma/\alpha$ & $7.854$ \\ 
$\re g$ & $-0.2234$ \\
$\im g$ & $\pm0.4469$ \\
$\alpha_5$ & $-0.1865$ \\
$\alpha_6$ & $-0.1116$ \\
$c_{13}$ & $-0.2405$ \\
$c_{23}$ & $-0.0959$ \\
\hdashline
$v_u^2 \kappa^2/\Lambda$~[eV] & $0.0191$ \\
\hdashline
$|\epsilon|\approx|\tau-i|$ & $0.1408$ \\
\hline
\end{tabular}
\hspace{0.2cm}
\begin{tabular}[t]{|c|c|}
\hline
\multicolumn{2}{|c|}{\texttt{Observables}} \\
\hline
\hline
$m_e/m_\mu$ & $0$ \\
$m_\mu/m_\tau$ & $0.0565$ \\
$r$ & $0.0299$\\
$\sin^2\theta_{12}$ & $0.304$ \\
$\sin^2\theta_{13}$ & $0.02219$ \\
$\sin^2\theta_{23}$ & $0.573$ \\
\hdashline
$\delta m^2~[10^{-5}~\text{eV}^2]$ & $7.42$ \\
$\Delta m^2~[10^{-3}~\text{eV}^2]$ & $2.480$ \\
\hline
\end{tabular}
\hspace{0.2cm}
\begin{tabular}[t]{|c|c|}
\hline
\multicolumn{2}{|c|}{\texttt{Predictions}} \\
\hline
\hline
$m_1$~[eV] & $0.0063$ \\
$m_2$~[eV] & $0.0107$ \\
$m_3$~[eV] & $0.0506$\\
$\delta/\pi$ & $\pm0.91$ \\
$\alpha_{21}/\pi$ & $\pm0.98$ \\
$\alpha_{31}/\pi$ & $\pm0.88$ \\
\hdashline
$|m_{ee}|$~[eV] & 0 \\
$\sum_i m_i$~[eV] & 0.0675 \\
\hdashline
Ordering & NO \\
\hdashline
$M_1/\Lambda$ & 0.353 \\
$M_2/\Lambda$ & 2.130 \\
$M_3/\Lambda$ & 2.483 \\
\hline
\end{tabular}
\caption{Second pair of best-fit points in a vicinity of $\tau = i$ found considering the K\"ahler potential in eqs.~\eqref{eq:KL_two_param}--\eqref{eq:X}. 
}
\label{tab:bfp2_model_2}
}
\end{table}

The inclusion of a non-minimal K\"ahler potential, even within the above restrictive assumptions, brings in several additional free parameters: $\alpha_{5,6}$, ${\tt Re}(c_{13,23})$ and ${\tt Im}(c_{13,23})$~\footnote{In our numerical analysis, we have set ${\tt Im}(c_{13,23})=0$.}. Adding them to $\beta/\alpha$, $\gamma/\alpha$, ${\tt Re}(g)$, ${\tt Im} (g)$, ${\tt Re}(\tau)$ and ${\tt Im}(\tau)$, we have a total of 12 dimensionless input parameters, more than the number of observables. Thus the focus of our analysis cannot be on predictability. 
Rather, we are interested in accounting for the mass hierarchies in terms of the $Z_4$ parameter $\epsilon$, in the context of a model reproducing all lepton masses and mixings. While the mass hierarchies alone can be easily accommodated without the need of hierarchical Lagrangian parameters, some degree of hierarchy turns out to be required by the need to fix the mixing parameters. Useful parameters to estimate such hierarchies in the K\"ahler potential are the 
condition numbers of eq.~(\ref{condnum}). 
To establish the possibility to reproduce all the relevant observables, and the role of $Z_4$ breaking in setting the mass hierarchies, we have selected several benchmark points with slightly different features. We show the results of two (pairs) of the benchmark points in Tables~\ref{tab:bfp_model_2} and \ref{tab:bfp2_model_2}. 
In all such benchmark points, all five dimensionless observables take exactly their 
experimental best-fit values (for the time being we set $m_e=0$). In addition, the model predicts a normal ordered neutrino mass spectrum and the values of the CPV phases. 
Interestingly, for both pairs of the benchmark points, 
the predicted value of $\delta$ (the one with minus sign) 
matches its experimental best-fit value.
Notice also the interesting result $\vert m_{ee}\vert =0$ which at the leading order can be seen as a simple consequence of the matrix patterns~\eqref{eq:seesaw at LO} and \eqref{me1}~\footnote{Given the column ordering in eq.~\eqref{me1}, $\vert m_{ee}\vert$ is given at the leading order by $\big(m_\nu ^{(0)}\big)_{33} = 0 + O(\epsilon)$.}. 
Finally, we report in the last column 
the masses $M_i$, $i=1,2,3$, of the heavy neutrinos in the units of $\Lambda$. Although the value of $\Lambda$ cannot be uniquely fixed, 
it can be estimated as (see the first column of the tables) 
$\Lambda \approx v_u^2\kappa^2/(0.02~\text{eV}) \approx 10^{15} \sin^2\beta$~GeV, where we have used $v_u = v \sin\beta$, with $v = 174$~GeV, and $\kappa \sim \mathcal{O}(1)$. This implies that for $\tan\beta \gtrsim 1$, 
the scale $\Lambda \gtrsim 5 \times 10^{14}$~GeV. 
Let us stress once again that in the considered model, $M_1 \sim \abs{\epsilon} \Lambda$, and thus, it is generated by a small departure of $\tau$ from $i$.

Our analysis shows that the mass hierarchies are indeed governed by $Z_4$ breaking, whereas, in general, K\"ahler corrections reflect on the lepton mass spectrum through $\mathcal{O}(1)$ changes.
For example, in the first pair of benchmark points (see Table~\ref{tab:bfp_model_2}), we verified numerically that K\"ahler corrections only affect the mass ratios by about a factor of 2~%
\footnote{For the minimal K\"ahler potential, 
\textit{i.e.} setting $\alpha_5 = \alpha_6 = c_{13} = c_{23} = 0$, 
we find $m_\mu/m_\tau = 0.0520$ and $r = 0.0637$, 
whereas the angles $\sin^2\theta_{12} = 0.228$, $\sin^2\theta_{13} = 0.03751$
and $\sin^2\theta_{23} = 0.256$ are far away from their experimental values.}; on the other hand, at these points, the K\"ahler metrics are by themselves somewhat hierarchical, as shown by the condition numbers $\kappa(K_L) \approx 12$ and $\kappa(K_{E^c}) \approx 16$. In the second pair of benchmark points shown in Table~\ref{tab:bfp2_model_2}, the hierarchies in the K\"ahler metrics are both reduced (the condition numbers are $\kappa(K_L) \approx 6$ and $\kappa(K_{E^c}) \approx 12$), and points with even milder hierarchies 
may potentially be found. 
These observations lead us to conclude that the deviations from the canonical K\"ahler metric 
present in the  best-fit points, have little to do with the mass spectrum hierarchies; 
rather, they are necessary in order to reproduce the correct PMNS mixing angles.

\begin{figure}[t]
\centering
\includegraphics[width=0.48\textwidth]{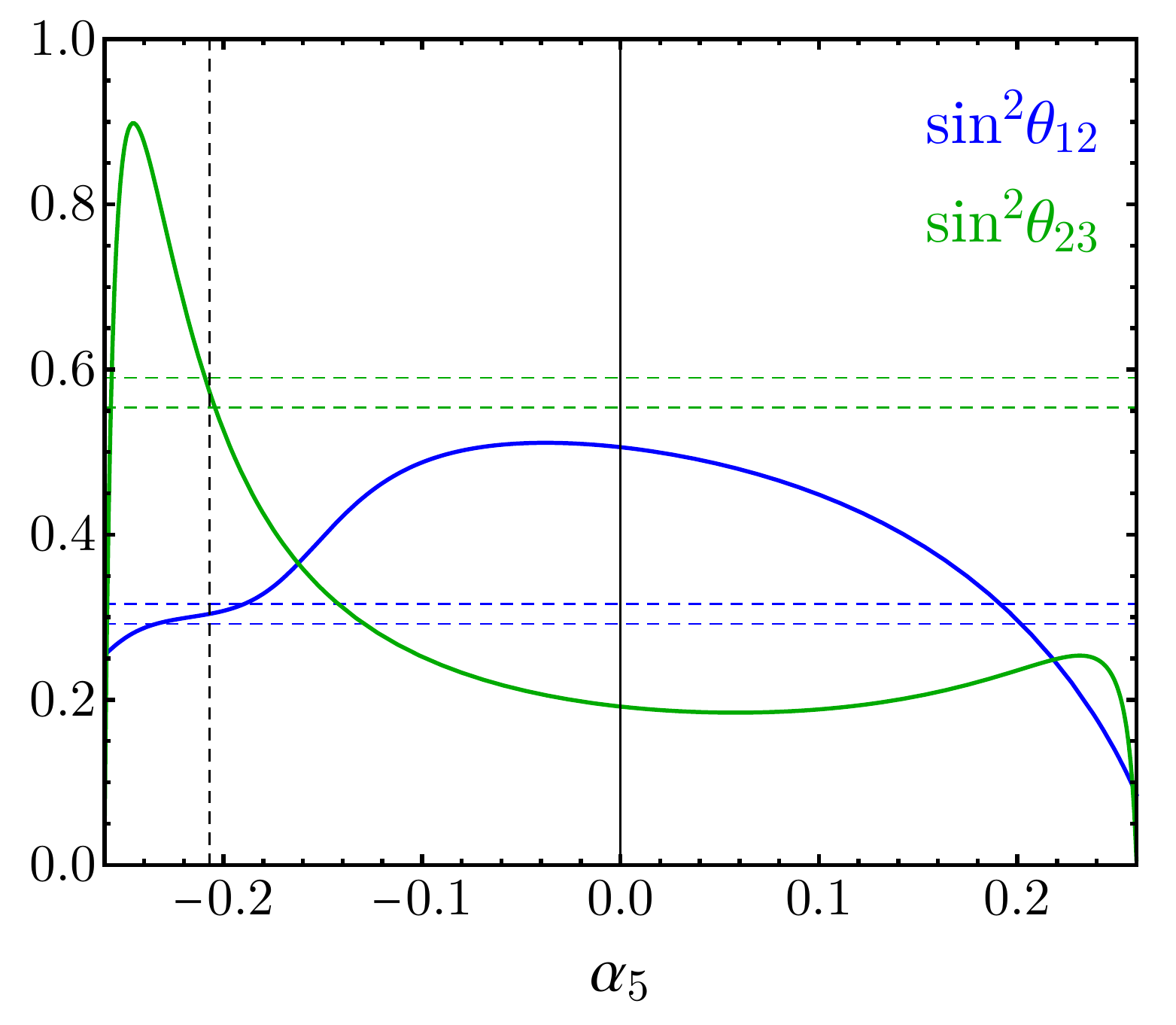}
\hspace{0.02\textwidth}
\includegraphics[width=0.48\textwidth]{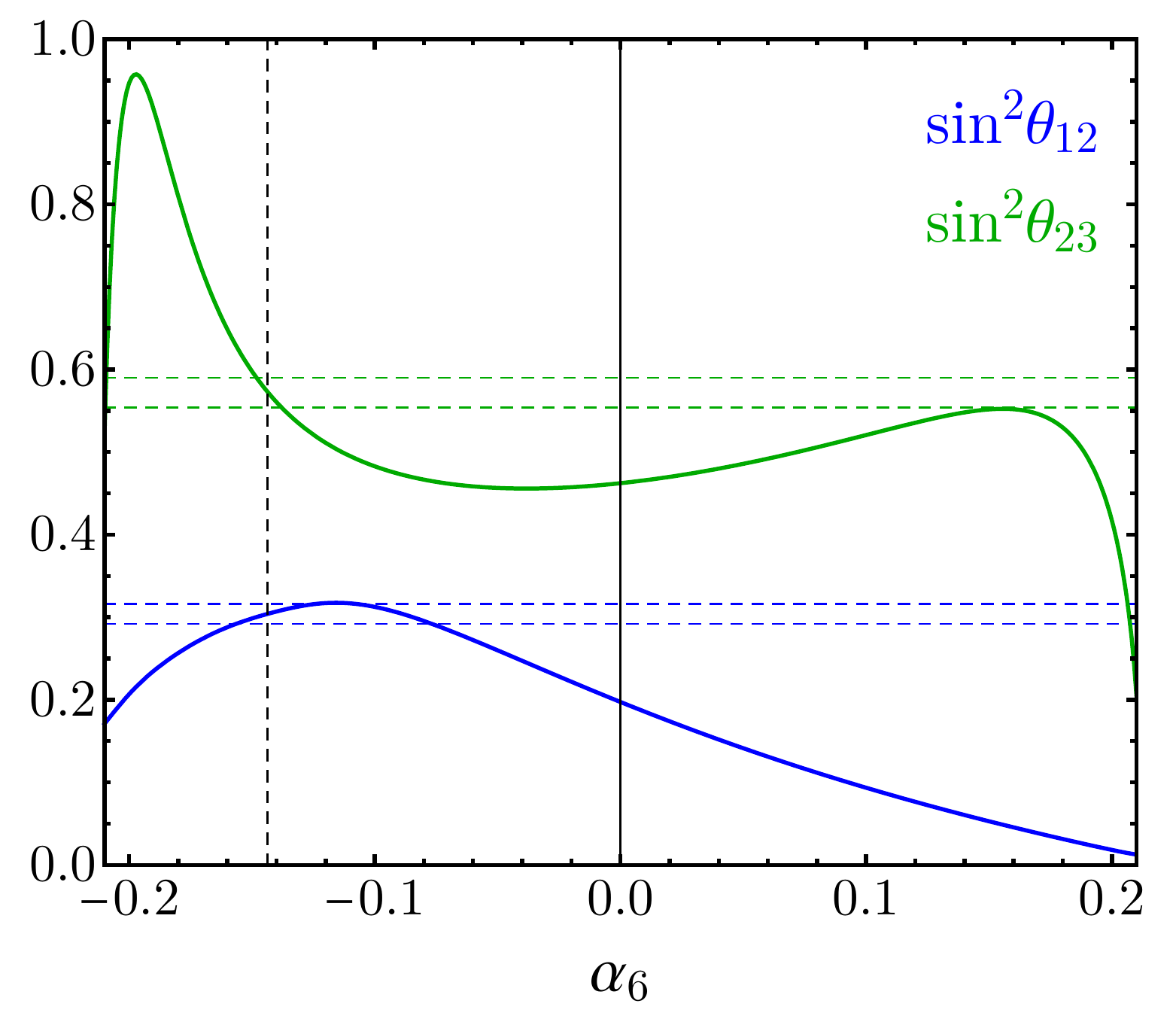}
\includegraphics[width=0.48\textwidth]{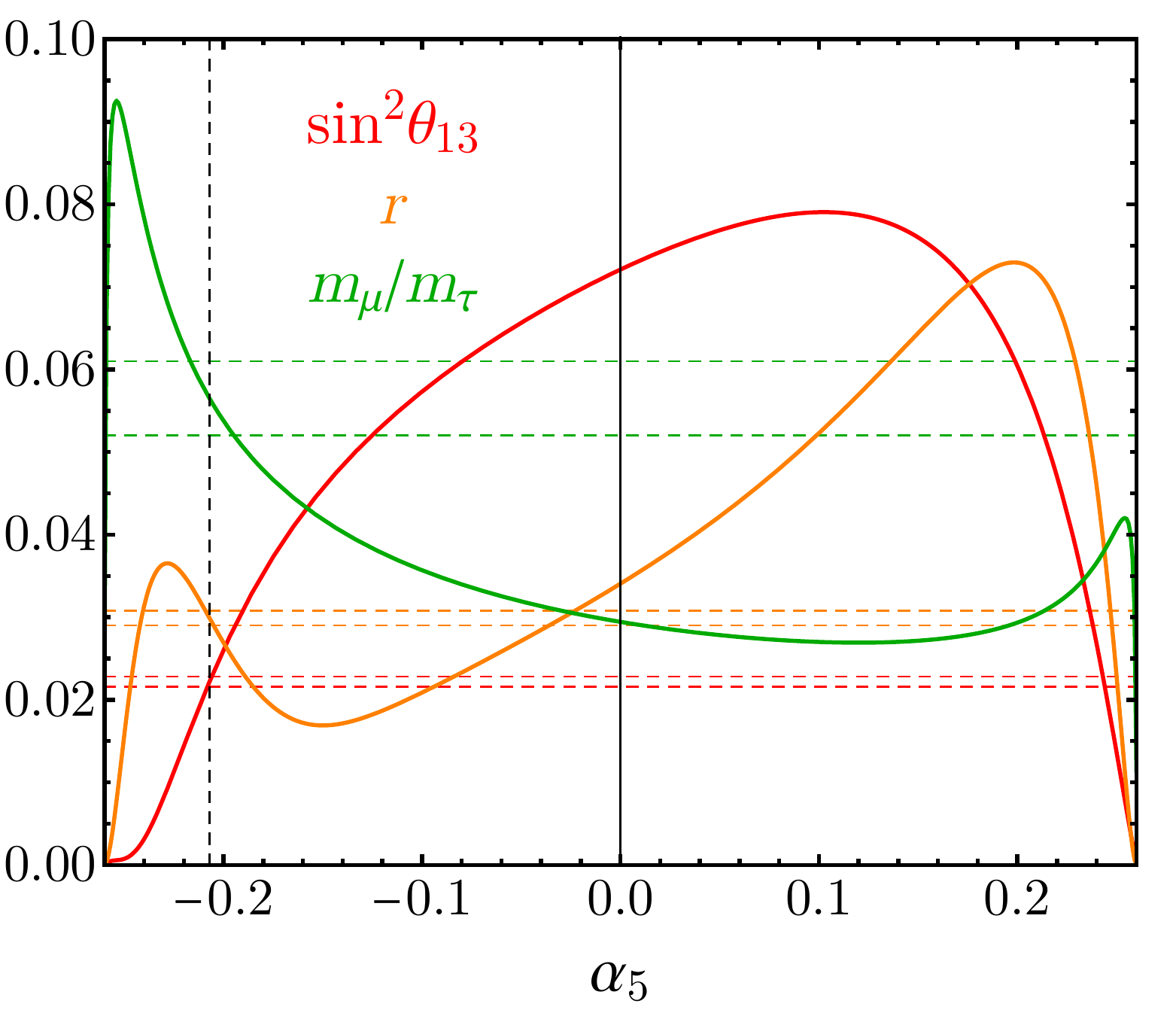}
\hspace{0.02\textwidth}
\includegraphics[width=0.48\textwidth]{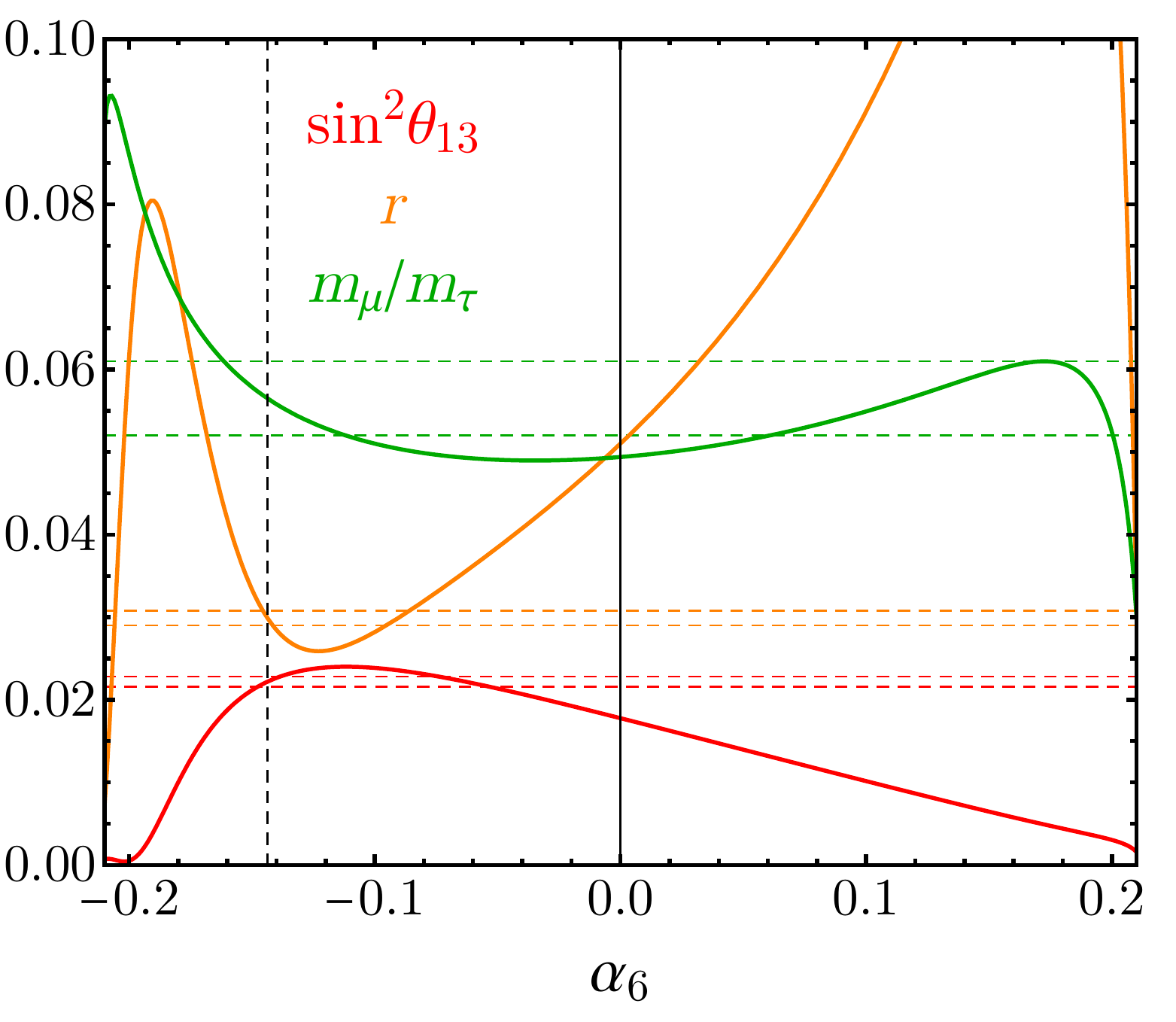}
\caption{Dependence of the mixing angles and two mass ratios on $\alpha_5$ (left) and $\alpha_6$ (right), fixing all other input parameters to their best-fit values from Table~\ref{tab:bfp_model_2}. The \textit{horizontal dashed lines} indicate the boundaries of the respective $1\sigma$ ranges from Table~\ref{tab:global_fit}. The \textit{vertical dashed line} in the left (right) panels stands for the best-fit value of $\alpha_5$ ($\alpha_6$) from Table~\ref{tab:bfp_model_2}.}
\label{fig:KL}
\end{figure}
%
%
%
\begin{figure}[t]
\centering
\includegraphics[width=0.48\textwidth]{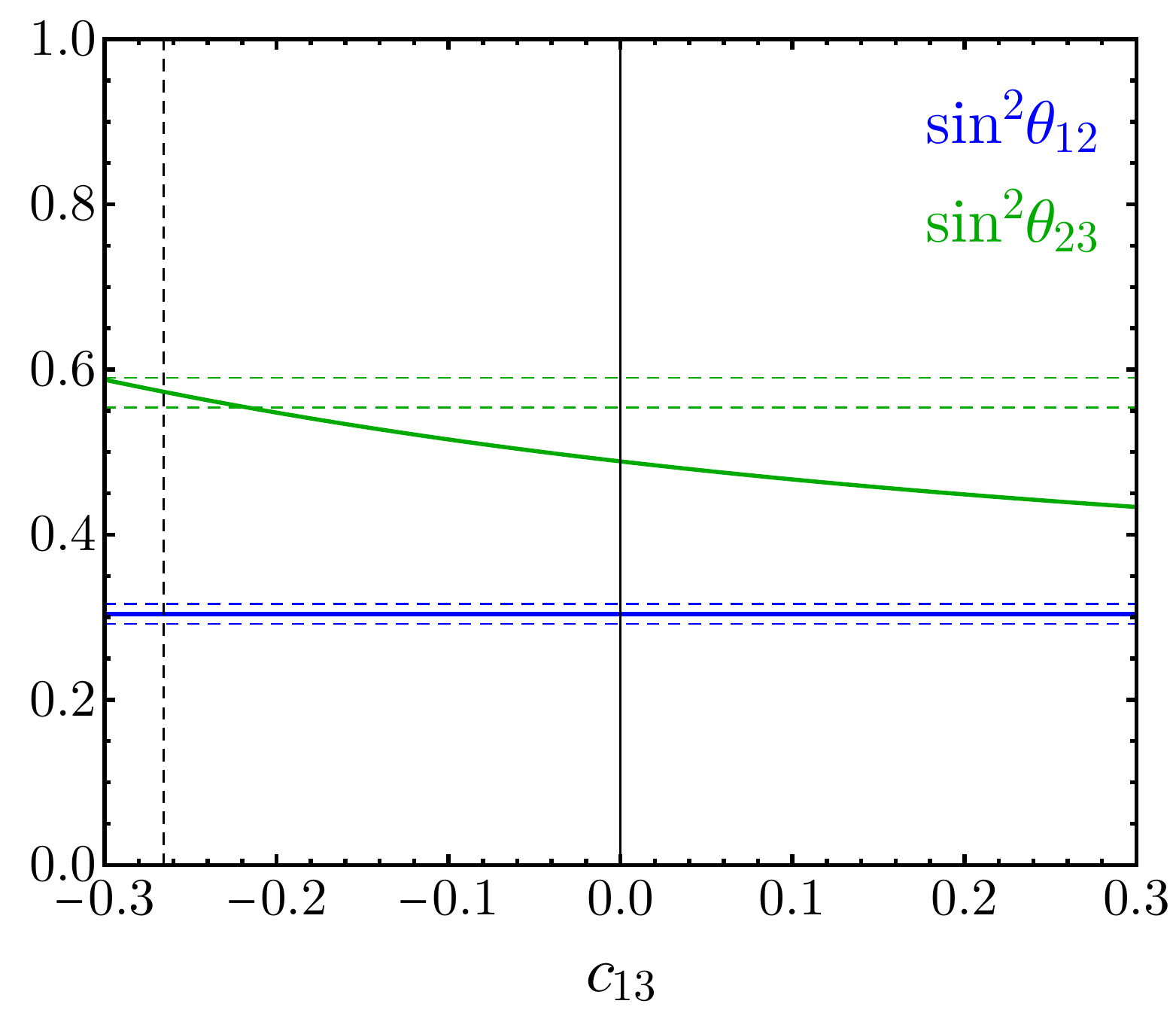}
\hspace{0.02\textwidth}
\includegraphics[width=0.48\textwidth]{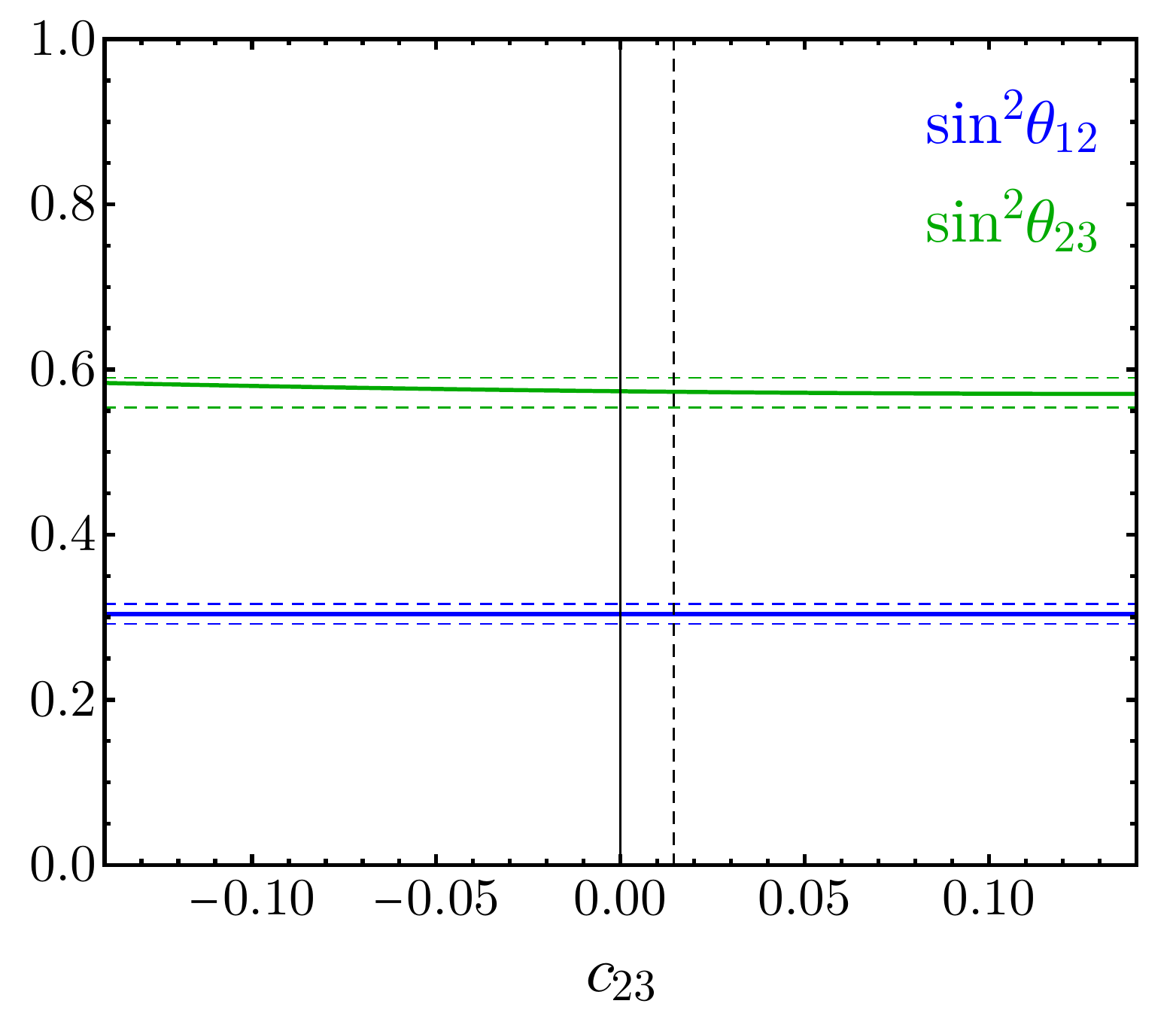}
\includegraphics[width=0.48\textwidth]{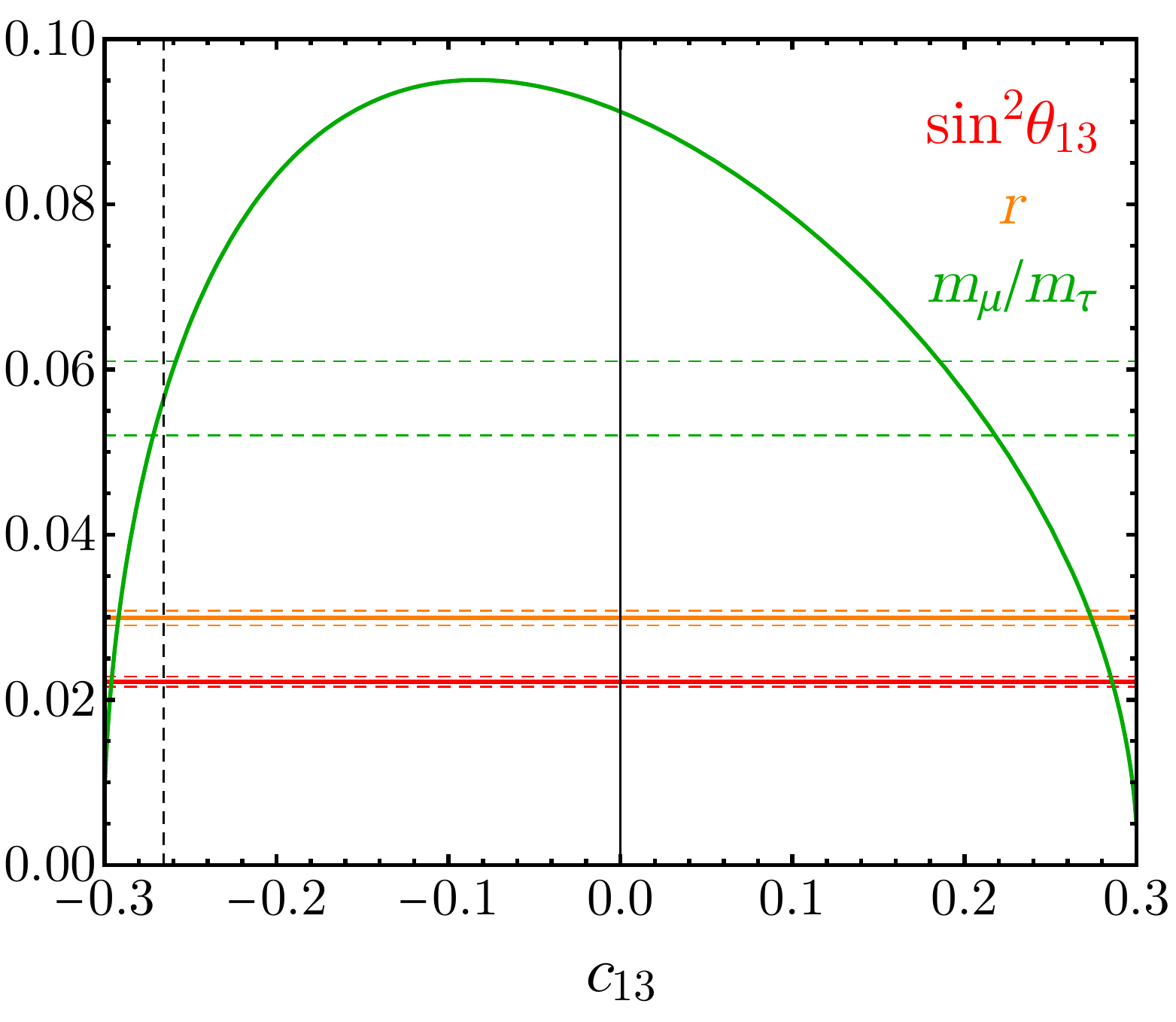}
\hspace{0.02\textwidth}
\includegraphics[width=0.48\textwidth]{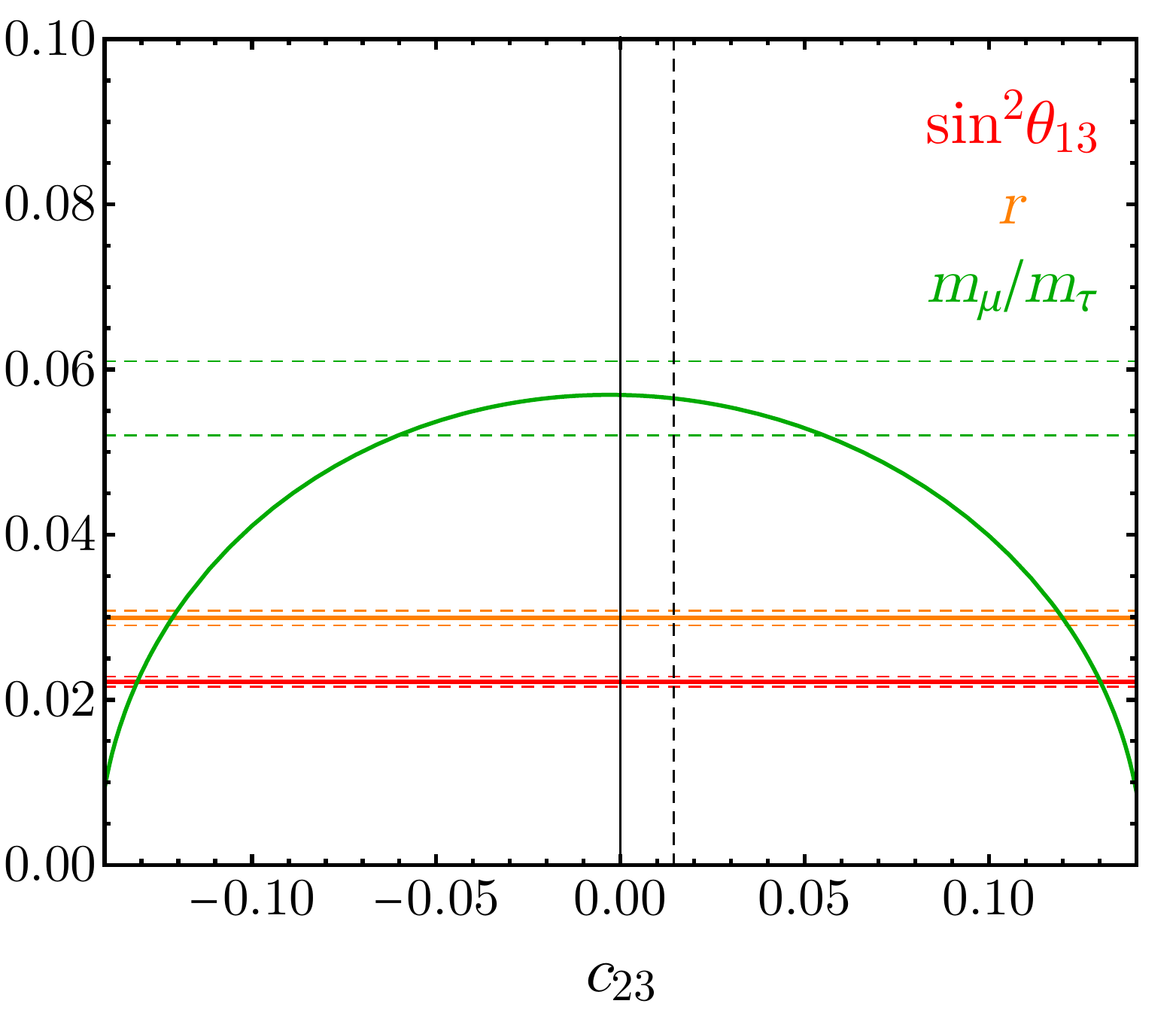}
\caption{Dependence of the mixing angles and two mass ratios on $c_{13}$ (left) and $c_{23}$ (right), fixing all other input parameters to their best-fit values from Table~\ref{tab:bfp_model_2}. The \textit{horizontal dashed lines} indicate the boundaries of the respective $1\sigma$ ranges from Table~\ref{tab:global_fit}. The \textit{vertical dashed line} in the left (right) panels stands for the best-fit value of $c_{13}$ ($c_{23}$) from Table~\ref{tab:bfp_model_2}.}
\label{fig:KEc}
\end{figure}
%
We have analysed more in detail the dependence of the fitted observables 
on the parameters of the K\"ahler potential. 
In Figs.~\ref{fig:KL} and \ref{fig:KEc}, we plot the values of 
the five dimensionless observables versus $\alpha_{5,6}$ and $c_{13,23}$, 
respectively~\footnote{For the values of $\alpha_{5,6}$ 
($c_{13,23}$) beyond the range displayed in the $x$-axis, 
the matrix $K_L$ ($K_{E^c}$) is not positive definite, 
and thus, the corresponding K\"ahler metric is well-defined 
only for the displayed range of $\alpha_{5,6}$ ($c_{13,23}$).}.
All other input parameters are fixed to their best-fit values 
as in Table~\ref{tab:bfp_model_2}.
We see that the parameters $\alpha_{5}$ and $\alpha_6$ strongly impact the predictions for 
the mixing angles and the two mass ratios, $r$ and $m_\mu/m_\tau$, 
whereas $c_{13}$ and $c_{23}$ in $K_{E^c}$ mainly affect the predictions 
for $\sin^2\theta_{23}$ and $m_\mu/m_\tau$. 

In conclusion, we see that a mass matrix of reduced rank at the self-dual point $\tau=i$ can 
explain the observed mass hierarchies in terms of $\ord{1}$ Lagrangian parameters. At the same time, at least in the model considered here, moderately hierarchical K\"ahler and superpotential parameters are needed to fix the mixing angle predictions.
Whether or not this is a general feature of this class of models is a question which definitely requires further investigation, but is beyond the scope of the present work. On the other hand, 
in order to get a fully realistic description of lepton masses we should still generate a non-vanishing electron mass, without perturbing too much the results achieved so far. We discuss this point in the next subsection.

\subsection{Generating $m_e \neq 0$}
\label{subsec:generating_me}
Both the models discussed above yield, by construction, $m_e = 0$. One can easily concoct mechanisms to generate the 
small electron mass without spoiling the  other predictions. We give below two examples, where $m_e$ is generated by supersymmetry breaking and by dimension six operators, respectively.

If supersymmetry is broken by some $F$-term, fermion masses get corrected by the second term of eq.~\eqref{eq:fermion_mass_mat}, which, as discussed below the same equation, scales as 
$m_\mathrm{SUSY}\, v/M$ for SM fermion masses (where $M$ is the SUSY breaking messenger scale). For instance, a K\"ahler interaction of the form:
\be
K\supset\frac{1}{\Lambda^{2}}\chi^{\dagger}E_{i}^{c}\left[a_{i}(\tau)H_{d}+b_{i}(\tau)\widetilde{H}_{u}\right]L+\text{h.c.}\,,
\ee
where the superfield $\chi$ gets a supersymmetry breaking expectation value $\langle\chi\rangle =F \theta^2$, gives a contribution to the charged lepton Yukawa matrix proportional to $F/\Lambda^2$, which in turn generically induces an electron Yukawa  coupling of the same order. 

As a second possibility,
one may generate $m_e \neq 0$ through the dimension six operator:
\be
(E_i ^c L H_d)(H_u H_d), \label{eq:dim6 superpotential}
\ee
whose Wilson coefficient should be a modular form of the appropriate weight. In order for this mechanism to work, we need to generalise the weight assignments in Table~\ref{tab:model_1}. We make the following requirements:
\begin{align}\label{eq:generalised_weights}
k_{L} & =1-k_{u}\,, \\
k_{E^{c}} & =3+k_{u}-k_{d}\,, \\
k_{u}+k_{d} & \neq 0 \,.
\end{align}
%
The first two conditions ensure that the superpotentials discussed in the previous subsections have weight zero; the last condition implies that the operator \eqref{eq:dim6 superpotential} has different weight from the corresponding renormalisable Yukawa term~%
\footnote{Curiously, the same condition can be exploited to make the Higgs $\mu$-term vanish at $\tau = i$. 
The Higgs $\mu$-term, being a $\Gamma _3$ singlet modular form of weight $k_u + k_d$,
vanishes by ~eq.~\eqref{es} at $\tau = i$ if $k_u + k_d \neq 0~(\texttt{mod}~4)$, since all $\Gamma _3$ singlets have $\rho(\widetilde S) = 1$.}, 
so that it couples to a functionally independent modular form multiplet 
(making the resulting charged lepton mass matrix of rank three). 
Such a mechanism thus generates $m_e \sim v_u v_d^2/\Lambda^2$, where $\Lambda$ is the scale at which the operator in eq.~\eqref{eq:dim6 superpotential} is generated.

While in some flavour models, the ratios $m_e/m_\tau$ and $m_\mu/m_\tau$ are associated to different powers of the same expansion parameter, we note that here the two ratios are associated to independent parameters.

\section{Conclusion}
\label{S5}
Supersymmetric modular invariant theories offer an attractive framework to address the flavour puzzle. The role of flavour symmetry is played by modular invariance, regarded as a discrete gauge symmetry, thus circumventing the obstruction concerning fundamental global symmetries.
The arbitrary symmetry breaking sector of the conventional models based on flavour symmetries is replaced by
the moduli space. Yukawa couplings become modular forms, severely restricted by the matter transformation properties. So far this framework has delivered interesting preliminary results
especially in the lepton sector, where neutrino masses and lepton mixing parameters can be efficiently
described in terms of a limited number of input parameters. 

Weak points in most of the existing constructions are the need of independent hierarchical parameters to describe
charged lepton masses, the reduced predictability caused by a general form of the K\"ahler potential, and
the absence of a reliable dynamical mechanism to determine the value of $\tau$ in the vacuum. 
As a matter of fact, in several models reproducing 
lepton masses and mixing parameters, the required value of $\tau$ is close to $i$, the self-dual point where
the generator $S$ of the modular group and CP (if the Lagrangian is CP invariant) are unbroken. 
A small departure of $\tau$ from $i$ suffices to generate sizeable CP-violating effects in the lepton sector.

For these reasons, we were led to analyse more in detail the vicinity of $\tau=i$. Our goal was to show that a small deviation
from the self-dual point can be responsible 
for the observed mass hierarchy $m_e\ll m_\mu\ll m_\tau$ and 
$\delta m^2 \ll \Delta m^2$. At $\tau=i$, the theory has an exact $Z_4$ symmetry, generated by the element $S$ of the
modular group. In the neighbourhood of $\tau=i$, the breaking of $Z_4$ can be fully described
by the (small) spurion $\epsilon\approx\tau-i$, that flips its sign under $Z_4$. 
We explained how to exploit this residual $Z_4$ symmetry in order to obtain lepton mass matrices having reduced rank at $\tau=i$. This can be easily done with a suitable assignment of modular weights and representations for matter fields. 
There is a twofold advantage in this strategy. First, mass ratios that are forced to vanish at $\tau=i$ by the $Z_4$ symmetry are expected to acquire small values $\propto |\epsilon|^n$  $(n>0)$ near the self-dual point.
Second, the reduced rank of the mass matrices can tame the contribution from a non-minimal K\"ahler potential,
provided the metrics of the matter fields do not display large hierarchies. 

To see whether this strategy can be successfully realised or not, 
we built a concrete model at level 3, where neutrinos get masses through
the type I seesaw mechanism.
The model predicts a normal mass ordering. The number of parameters
exceeds the number of fitted observables and we cannot claim predictability. However, with $\tau$ being near $i$, 
mass ratios and mixing angles are
reproduced with input parameters nearly of the same order of magnitude
and matter kinetic terms 
display only a moderate hierarchy. 
We saw that the 
main contribution to the 
mass hierarchy can be induced by
the singular mass matrix at the $Z_4$ symmetric point. In the model we considered, the K\"ahler potential and the other Lagrangian parameters are crucial in order to correctly reproduce the values of the mixing angles.
While the $Z_4$ symmetry plays a fundamental role in all our discussion, we notice that
our model could not have been realised in the context of a $Z_4$ flavour symmetry alone. In particular, the  electron mass vanishes in the models we have considered due to the correlations among generic $Z_4$-invariant operators provided by the underlying modular invariance. Also the leading order values of the mixing angles are dictated by $Z_4$.
We have discussed possible sources of a non-vanishing electron mass.
While the models we formulated have clearly room for improvement, we consider them as a good starting point 
to naturally accommodate the observed fermion mass hierarchies within a modular invariant framework. 

\clearpage
\textit{Note Added.} Few weeks after completion of this work, 
ref.~\cite{Novichkov:2021evw} appeared on the arXiv, in which the authors performed
a systematic study of all fixed points, 
$\tau_S = i$, $\tau_{ST} = e^{2 \pi i / 3}$, and $\tau_T = i \infty$, assuming a minimal form of the K\"ahler potential. 
The points $\tau _{ST}$ and $\tau _T$ enjoy residual $Z_3 \times Z_2$ and $Z_N \times Z_2$ symmetries, 
respectively (where $N$ is the level of modular forms used in the construction, $N=3$ in our work).
It is found, in particular, that fermion mass hierarchies crucially depend on 
the decomposition of field representations under the residual symmetry group.

\section*{Acknowledgements}
This project has received support in part by the MIUR-PRIN project 2015P5SBHT 003 ``Search for the Fundamental Laws and Constituents'' and by the European Union's Horizon 2020 research and innovation programme under the Marie Sk\l{}odowska-Curie grant agreement N$^\circ$~860881-HIDDeN.
The research of F.~F.~was supported in part by the INFN. 
The work of A.~T. was partially supported by the FEDER/MCIyU-AEI grant FPA2017-84543-P and by the “Generalitat Valenciana” under grant PROMETEO/2019/087.

\appendix
\section{K\"ahler potential quadratic in $Y^{(2)}$}
\label{app:Kahler}
In the real basis for the $\Gamma_3$ generators $S$ and $T$ 
in the 3-dimensional representation we employ in this work, 
$L^\ast$ and $Y^{(2)\ast}$ transform as triplets, i.e.\ 
$L^\ast \to \rho_\3(\tilde\gamma)L^\ast$ and 
$Y^{(2)\ast} \to \rho_\3(\tilde\gamma)Y^{(2)\ast}$. 
Thus, we can contract first $L^\ast$ with $L$ and $Y^{(2)\ast}$ with $Y^{(2)}$, 
and after that perform contractions of the obtained multiplets. 
Proceeding in this way, we obtain:
\begin{align}
 \left(L^\ast L\right)_\1 &= L^\dagger L\,, 
 \qquad
 \left(L^\ast L\right)_\op = L^\dagger M_\op L\,,
 \qquad
 \left(L^\ast L\right)_\opp = L^\dagger M_\opp L\,, \\
 \left(L^\ast L\right)_{\3_S} &=\begin{pmatrix}
 L^\dagger M_{\3_S}^{(1)} L \\
 L^\dagger M_{\3_S}^{(2)} L \\
 L^\dagger M_{\3_S}^{(3)} L
 \end{pmatrix}, 
 \qquad
 \left(L^\ast L\right)_{\3_A} =\begin{pmatrix}
 L^\dagger M_{\3_A}^{(1)} L \\
 L^\dagger M_{\3_A}^{(2)} L \\
 L^\dagger M_{\3_A}^{(3)} L
 \end{pmatrix},
\end{align}
%
with the matrices $M_\r$ being
\begin{align}
 M_\op &= \begin{pmatrix}
 1 & 0 & 0 \\
 0 & \omega^2 & 0 \\
 0 & 0 & \omega 
 \end{pmatrix}, 
 \qquad
 M_\opp = \begin{pmatrix}
 1 & 0 & 0 \\ 
 0 & \omega & 0 \\
 0 & 0 & \omega^2 
 \end{pmatrix}, 
 \label{eq:M1p_M1pp} \\
 M_{\3_S}^{(1)} &= \begin{pmatrix}
 0 & 0 & 0 \\
 0 & 0 & 1 \\
 0 & 1 & 0 
 \end{pmatrix}, 
 \qquad
 M_{\3_S}^{(2)} = \begin{pmatrix}
 0 & 0 & 1 \\
 0 & 0 & 0 \\
 1 & 0 & 0 
 \end{pmatrix}, 
 \qquad
 M_{\3_S}^{(3)} = \begin{pmatrix}
 0 & 1 & 0 \\
 1 & 0 & 0 \\
 0 & 0 & 0 
 \end{pmatrix}, \\
 M_{\3_A}^{(1)} &= \begin{pmatrix}
 0 & 0 & 0 \\
 0 & 0 & 1 \\
 0 & -1 & 0 
 \end{pmatrix}, 
 \qquad
 M_{\3_A}^{(2)} = \begin{pmatrix}
 0 & 0 & -1 \\
 0 & 0 & 0 \\
 1 & 0 & 0 
 \end{pmatrix}, 
 \qquad
 M_{\3_A}^{(3)} = \begin{pmatrix}
 0 & 1 & 0 \\
 -1 & 0 & 0 \\
 0 & 0 & 0 
 \end{pmatrix},
\end{align}
%
and $\omega = e^{2\pi i/3}$. 
The same equations hold for $\left(Y^{(2)\ast} Y^{(2)}\right)_\r$. 
Taking further invariant contractions of the obtained multiplets, 
we find
\clearpage
\begin{align}
K_L = 
(2\im\tau)^{-k_L} \mathbb{1} \phantom{{}=}& \nonumber \\
{}+{} (2\im\tau)^{k_Y-k_L} \bigg\lbrace&
\alpha_1 Y^{(2)\dagger} Y^{(2)} \mathbb{1} 
+ \alpha_2 \left[\left(Y^{(2)\dagger} M_\opp Y^{(2)}\right) M_\op 
+ (Y^{(2)\dagger} M_\op Y^{(2)}) M_\opp\right] \nonumber \\
&+ \alpha_3\, i \left[\left(Y^{(2)\dagger} M_\opp Y^{(2)}\right) M_\op 
- \left(Y^{(2)\dagger} M_\op Y^{(2)}\right) M_\opp\right] \nonumber\\
&+ \alpha_4 \sum_{n=1}^{3} \left(Y^{(2)\dagger} M_{\3_S}^{(n)} Y^{(2)}\right) M_{\3_S}^{(n)} 
+ \alpha_5 \sum_{n=1}^{3} \left(Y^{(2)\dagger} M_{\3_A}^{(n)} Y^{(2)}\right) M_{\3_A}^{(n)} \nonumber \\
&+ \alpha_6 \sum_{n=1}^{3} i \left(Y^{(2)\dagger} M_{\3_A}^{(n)} Y^{(2)}\right) M_{\3_S}^{(n)} 
+ \alpha_7 \sum_{n=1}^{3} i \left(Y^{(2)\dagger} M_{\3_S}^{(n)} Y^{(2)}\right) M_{\3_A}^{(n)} 
\bigg\rbrace\,,
\label{eq:KL}
\end{align}
%
where $\alpha_j$, $j=1\,,\dots\,,7$, are real coefficients, which accompany 
hermitian matrices. (We have used the fact that $M_\op^\dagger = M_\opp$, 
$M_{\3_S}^{(n)\dagger} = M_{\3_S}^{(n)}$, and
$M_{\3_A}^{(n)\dagger} = -M_{\3_A}^{(n)}$.)

One of our assumptions is that the canonical form of $K_L$ 
is restored at $\im\tau \to \infty$. 
The $q$-expansions of $Y_i$ in the real basis read:
\begin{align}
Y_1(\tau) &= \frac{1}{\sqrt{3}}\left(1 - 6q^{1/3} - 18q^{2/3} + 12q + \dots\right), \\
Y_2(\tau) &= \frac{1}{\sqrt{3}}\left(1 - 6\omega q^{1/3} - 18\omega^2 q^{2/3} + 12q + \dots\right), \\
Y_3(\tau) &= \frac{1}{\sqrt{3}}\left(1 - 6\omega^2 q^{1/3} - 18\omega q^{2/3} + 12q + \dots\right),
\end{align}
%
where $q=e^{2\pi i \tau}$. 
Thus, at $\im\tau \to \infty$
\begin{equation}
Y_1 = Y_2 = Y_3 \sim \frac{1}{\sqrt{3}}\,,
\end{equation}
%
and $K_L$ has the following form:
\begin{equation}
K_L \sim \frac{1}{2\im\tau} \mathbb{1} + \frac{2}{3}\im\tau 
\begin{pmatrix}
\frac{3}{2}\alpha_1 & \alpha_4+i\alpha_7 & \alpha_4-i\alpha_7 \\
 \alpha_4-i\alpha_7 & \frac{3}{2}\alpha_1 & \alpha_4+i\alpha_7 \\
\alpha_4+i\alpha_7 & \alpha_4-i\alpha_7 & \frac{3}{2}\alpha_1
\end{pmatrix},
\label{eq:KLinf}
\end{equation}
%
where we have used $k_L=1$ and $k_Y = 2$.
To satisfy our assumption of the asymptotic behaviour of $K_L$, 
the coefficients $\alpha_1 = \alpha_4 = \alpha_7 = 0$. 
Thus, the number of free parameters in $K_L$ is reduced from seven to four. 
Then, the elements of $K_L$ from eq.~\eqref{eq:KL} read:
\begin{align}
\left(K_L\right)_{11} &= \frac{1}{2\im\tau} + 2\im\tau 
\left[ 2\alpha_2\abs{Y_1}^2 - 
\left(\alpha_2+\sqrt{3}\alpha_3\right)\abs{Y_2}^2 - 
\left(\alpha_2-\sqrt{3}\alpha_3\right)\abs{Y_3}^2\right], 
\label{eq:KL11}\\
\left(K_L\right)_{22} & = \frac{1}{2\im\tau} + 2\im\tau 
\left[-\left(\alpha_2-\sqrt{3}\alpha_3\right)\abs{Y_1}^2 + 
2\alpha_2\abs{Y_2}^2 - 
\left(\alpha_2+\sqrt{3}\alpha_3\right)\abs{Y_3}^2\right], \\
\left(K_L\right)_{33} & = \frac{1}{2\im\tau} + 2\im\tau 
\left[-\left(\alpha_2+\sqrt{3}\alpha_3\right)\abs{Y_1}^2 - 
\left(\alpha_2-\sqrt{3}\alpha_3\right)\abs{Y_2}^2 + 
2\alpha_2\abs{Y_3}^2\right], \\
\left(K_L\right)_{12} & = 2\im\tau 
\left(\alpha_5+i\alpha_6\right) \left[Y_1^\ast Y_2 - Y_1 Y_2^\ast\right], \\
\left(K_L\right)_{13} & = 2\im\tau 
\left(\alpha_5-i\alpha_6\right) \left[Y_1^\ast Y_3 - Y_1 Y_3^\ast\right], \\
\left(K_L\right)_{23} &= 2\im\tau 
\left(\alpha_5+i\alpha_6\right) \left[Y_2^\ast Y_3 - Y_2 Y_3^\ast\right]. 
\label{eq:KL23}
\end{align}
%
For the sake of simplicity, we set further $\alpha_2=\alpha_3=0$.
In this case, the diagonal entries of $K_L$ are not affected by the contributions 
containing modular forms, on the contrary to the off-diagonal elements. 
Thereby, we arrive at the form of $K_L$ 
in eqs.~\eqref{eq:KL_two_param} and \eqref{eq:Xij}. 

What concerns $K_{E^c}$, with the assignment of representations and weights 
given in Table~\ref{tab:model_1}, the most general K\"ahler potential 
quadratic in $Y^{(2)}$ reads 
\begin{align}
K_{E^c} &= \frac{1}{8 \left(\im\tau\right)^3} \begin{pmatrix}
c_{11}^{0} & c_{12}^{0} & 0 \\
c_{12}^{0\ast} & c_{22}^{0} & 0 \\
0 & 0 & c_{33}^{0}
\end{pmatrix} \nonumber \\
&+\frac{1}{2\im\tau} \begin{pmatrix}
c_{11} Y^{(2)\dagger} Y^{(2)} & c_{12} Y^{(2)\dagger} Y^{(2)} & c_{13} Y^{(2)\dagger} M_\opp Y^{(2)} \\
c_{12}^\ast Y^{(2)\dagger} Y^{(2)} & c_{22} Y^{(2)\dagger} Y^{(2)} & c_{23} Y^{(2)\dagger} M_\opp Y^{(2)} \\
c_{13}^\ast Y^{(2)\dagger} M_\op Y^{(2)} & c_{23}^\ast Y^{(2)\dagger} M_\op Y^{(2)} & c_{33} Y^{(2)\dagger} Y^{(2)}
\end{pmatrix}\,,
\end{align}
%
with $c_{ii}^{(0)}$ being real and $c_{ij}^{(0)}$, $i \neq j$, complex coefficients.

Taking into account that at $\im\tau \to \infty$, the invariant combination 
$Y^{(2)\dagger} Y^{(2)} \sim 1$, whereas 
$X \equiv Y^{(2)\dagger} M_\opp Y^{(2)}$ and 
$X^\ast = Y^{(2)\dagger} M_\op Y^{(2)}$ decay exponentially, we find 
\begin{equation}
K_{E^c} \sim \frac{1}{8 \left(\im\tau\right)^3} \begin{pmatrix}
c_{11}^{0} & c_{12}^{0} & 0 \\
c_{12}^{0\ast} & c_{22}^{0} & 0 \\
0 & 0 & c_{33}^{0}
\end{pmatrix}
+
\frac{1}{2 \im\tau} \begin{pmatrix}
c_{11} & c_{12} & 0 \\
c_{12}^\ast & c_{22} & 0 \\
0 & 0 & c_{33}
\end{pmatrix}.
\end{equation}
%
In order to restore the canonical form of $K_{E^c}$ in the considered limit, $c_{12}^{0} = c_{12} = 0$. 
Furthermore, we set $c_{ii} = 0$ for simplicity.
Finally, we can always make $c_{ii}^{0} = 1$ by independent rescalings 
of $E_i^c$, $i=1,2,3$.
Thus, we recover $K_{E^c}$ given by eqs.~\eqref{eq:KEc_two_param} and \eqref{eq:X}.

\bibliographystyle{utphys}
\bibliography{paper}

\end{document}